# Properties of HPK UFSD after neutron irradiation up to 6e15 n/cm$^2$

Z. Galloway, V. Fadeyev, P. Freeman, E. Gkougkousis[2], C. Gee, B. Gruey, C.A. Labitan, Z. Luce, F. McKinney-Martinez, H. F.-W. Sadrozinski[1], A. Seiden, E. Spencer, M. Wilder, N. Woods, A. Zatserklyaniy, Y. Zhao

*SCIPP, Univ. of California Santa Cruz, CA 95064, USA*

N. Cartiglia, M. Ferrero, M. Mandurrino, A. Staiano, V. Sola

*INFN, Torino, Italy*

R. Arcidiacono

*Università del Piemonte Orientale and INFN, Italy*

V. Cindro, G. Kramberger, I. Mandić, M. Mikuž, M. Zavrtanik

*Jožef Stefan institute and Department of Physics, University of Ljubljana, Ljubljana, Slovenia*

*Abstract*– In this paper we report results from a neutron irradiation campaign of Ultra-Fast Silicon Detectors (UFSD) with fluences of 1e14, 3e14, 6e14, 1e15, 3e15 and 6e15 neq/cm$^2$. The UFSD used in this study are circular 50 μm thick Low-Gain Avalanche Detectors (LGAD), with a 1.0 mm diameter active area. Hamamatsu Photonics (HPK), Japan, produced the UFSD with pre-irradiation internal gain in the range 5-70 depending on the bias voltage. The sensors were tested pre-irradiation and post-irradiation with minimum ionizing particles (MIPs) from a $^{90}$Sr β-source. The leakage current, internal gain and the timing resolution were measured as a function of bias voltage at -20 $^o$C and -30 $^o$C. The timing resolution of each device under test was extracted from the time difference with a second calibrated UFSD in coincidence, using the constant fraction discriminator (CFD) method for both. The dependence of the gain upon the irradiation fluence is consistent with the acceptor removal mechanism; at -20 $^o$C the highest gain decreases from 70 before radiation to 2 after a fluence of 6e15 n/cm$^2$. Consequently, the timing resolution was found to deteriorate from 20 ps to 50 ps. The results indicate that the most accurate time resolution is obtained varying with fluence the CFD value used to determine the time of arrival, from 0.1 for pre-irradiated sensors to 0.6 at the highest fluence. Key changes to the pulse shape induced by irradiation, i.e. (i) the contribution of charge multiplication not limited to the gain layer zone, (ii) the shortening of the rise time and (iii) the reduced pulse height, were compared with the WF2 simulation program and were found to be in agreement.

1 Corresponding author: hartmut@ucsc.edu, telephone (831) 459 4670, FAX (831) 459 5777

2 Now at IFAE, Barcelona, Spain

## 1 INTRODUCTION

We are developing a new type of silicon detector, the so-called ultra-fast silicon detector (UFSD) that would establish a new paradigm for space-time particle tracking [1]. The UFSD is a single device that ultimately will measure with high precision concurrently the space (~10 μm) and time (~10 ps) coordinates of a particle.
UFSD are thin pixelated n-on-p silicon sensors based on the Low-Gain Avalanche Detector (LGAD) design [2][3][4] developed by the Centro Nacional de Microelectrónica (CNM) Barcelona, in part as a RD50 Common Project [5]. The sensor exhibits moderate internal gain (~5-70) due to a highly doped $p^+$ region just below the n-type implants. In [6] a time resolution below 35 ps was achieved in a beam test with un-irradiated 45 μm thick UFSD fabricated by CNM. This result complemented previous measurements on thicker sensors in beam tests and laboratory reported in [7][8]. These sets of measurements, taken with LGAD of different thicknesses, agreed well with the predictions of the simulation program Weightfield2 (WF2) [9].
First applications of UFSD are envisioned in the upgrades of the ATLAS and CMS experiments at the High-Luminosity Large Hadron Collider (HL-LHC [10]) as reviewed in [11]. In both experiments, the UFSD would be of moderate segmentation (a few $mm^2$) and will face challenging radiation requirements (fluences up to several $10^{15}$ $neq/cm^2$ and several hundred of MRad). Results on irradiated CNM LGAD 300 μm, 75 μm and 45 μm thick sensors are presented in [12][13][14], where the timing resolution has been shown to deteriorate with fluence due to the decreasing value of the gain. This effect is caused by the "acceptor removal mechanism" [15] that decreases the active dopant in the gain layer.
In this paper, we report on the results of the irradiation campaign of UFSD produced by Hamamatsu Photonics, Japan (HPK). In Section 2 we will describe the characteristics of the 50 μm UFSD manufactured by HPK followed in Section 3 by a short description of the neutron irradiation facility. In Section 4, a description of the experimental set-up is presented, including the readout electronics and the laboratory $^{90}Sr$ β-source used for charge collection studies mentioned in [6]. In Section 5, we will describe the data analysis including the extraction of the gain, timing resolution and pulse characteristics and in Section 6 the results on bias dependence of leakage current, charge collection and gain, pulse characteristics and timing resolution for a range of neutron fluences will be presented. Section 7 contains the comparison of results with predictions by WF2.

## 2 PRE-IRRADIATION PROPERTIES OF HPK UFSD

The UFSD were manufactured by HPK on 6 inch silicon wafers of 150 μm total thickness with a 50 μm or 80 μm thick high resistivity float zone (FZ) active layer. HPK run #ECX20840 contains a variety of pad structures, notably single pads of area ~0.8 $mm^2$ and 2x2 pad arrays with each pad 3x3 $mm^2$. The study presented in this paper uses 50 μm thick single pads, which were produced in four "gain splits" (named "A" to "D"), identical in the mask design but with a different $p^+$ doping profile of the gain layer. These gain splits are used to study the doping profile parameters that optimize the time resolution before and after irradiation. The C-V measurements indicate that the difference between doping concentrations of adjacent splits is about 4% [16]. The leakage currents are very low before breakdown, less than 100 pA.

The gain was measured using the read-out board described in Section 3 and β-particles from a $^{90}$Sr source. As shown later, the gain is computed by comparing the collected charge to the charge expected from a MIP in an equivalent sensor without gain. Using [17] this is estimated to be 0.51 fC for 50 µm thick un-irradiated sensors, while it was calculated at each fluence step using WF2 [9].

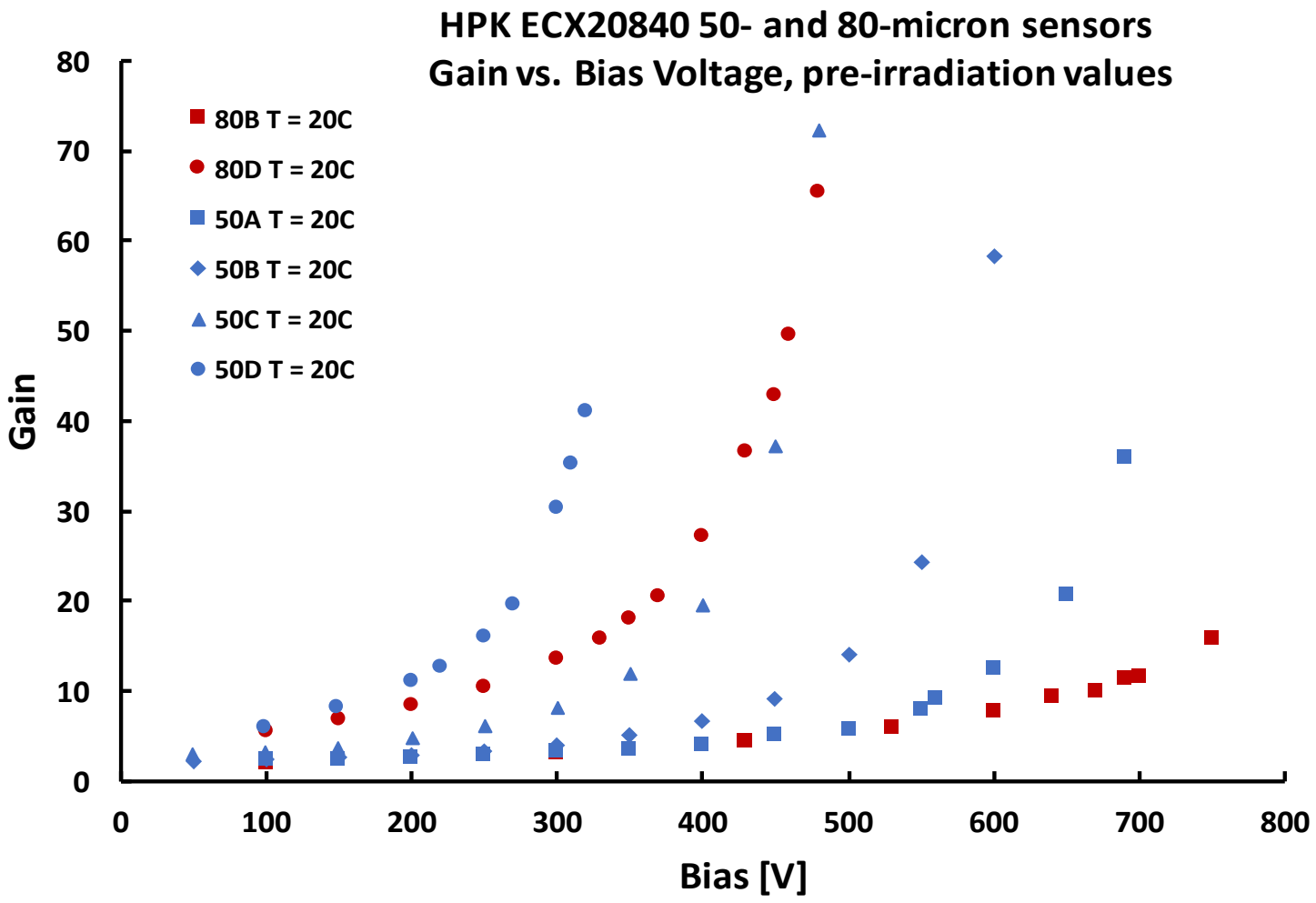

Figure 1 Gain vs. Bias voltage for the HPK 50 µm and 80 µm thick LGAD at +20 $^{o}$C. The four different sensors A-D reflect the four different doping profiles of the multiplication layer. The gain has a common systematic scale error of 20%.

Figure 1 shows the measured gain as a function of bias voltage for 50 µm thick single pads from the four splits A-D, and for two pads of 80 µm thickness for splits B and D. Figure 1 shows that the split D has the highest gain, and therefore the highest initial doping concentration, while split A has the lowest.

The results presented in this study are obtained using sensors of type 50D. Sensors with the highest initial doping concentration were chosen because irradiated CNM LGAD showed that sensors with higher initial doping concentration retained higher acceptor concentration after irradiation than those with lower initial concentration [12][13][14]. For the irradiation study, six samples of 50D were selected. All were tested in a probe station before irradiation at 20 $^{o}$C to assess their breakdown voltages (VBD = 350 +-20V) and to measure their C-V characteristics. Additional measurement on HPK sensors from the same production but not part of this study can be found in [16][18][19]

## 3 Neutron Irradiation

The LGAD were irradiated without bias in the TRIGA research reactor of the Institut Jozef Stefan in Ljubljana, which has been used successfully in the past decades to support sensor development [20]. The neutron spectrum and flux are well known and the fluence is quoted in 1 MeV equivalent neutrons per cm$^2$ (neq/cm$^2$ or shortened n/cm$^2$). For each of the following fluence points 1e14, 3e14, 6e14, 1e15, 3e15 and 6e15 neq/cm$^2$ one LGAD was irradiated.

After irradiation, the devices were annealed for 80 min at 60 $^{o}$C following standard RD50 practices to account for the expected long-term annealing during operation. Capacitance-voltage scans were identical before and after annealing. The devices were kept in cold storage at -20 $^{o}$C when they weren't being tested.

## 4 Experimental set-up

### *Detector Readout*

For charge collection studies the UFSD were mounted on a 10x10 $cm^2$ read-out board developed at the University of California Santa Cruz (UCSC). It has been used in previous beam tests and is described in detail in [6]. The first inverting amplifier on the UCSC board is followed by an external commercial 20dB amplifier (a Mini-Circuits TB-409-52+) to give a total trans-impedance of 4700 Ω. This value of the trans-impedance, relating the detector current to the measured voltage signal, has been simulated with SPICE and cross checked with measurements with different signal sources (β-particles, laser, minimum ionizing particles with slow amplifiers) and electronics (integrating and fast read-out). From these studies, a systematic scale error of 20% common to all charge collection measurements has been estimated.

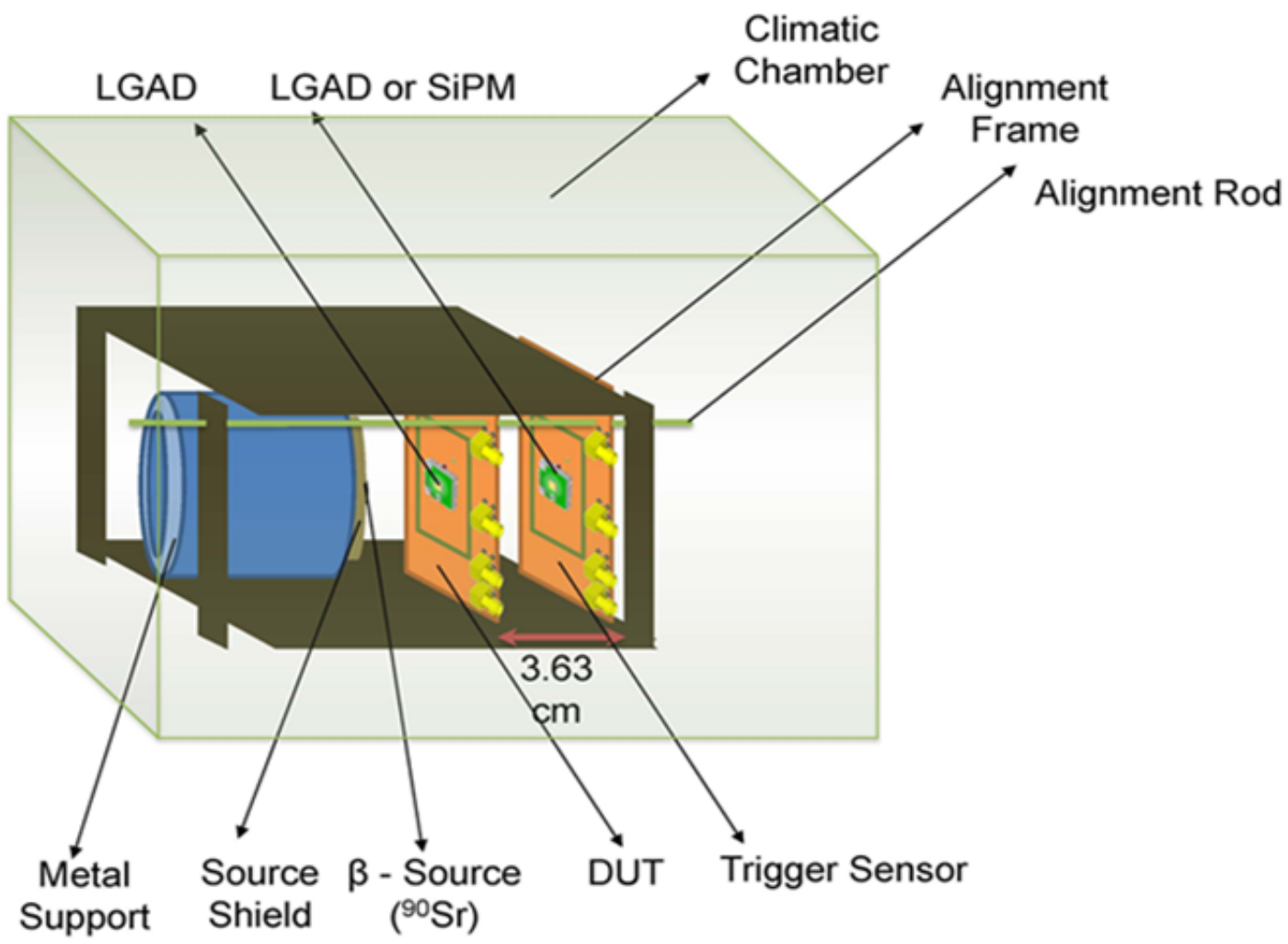

Figure 2. The laboratory setup of the β-telescope consisting of alignment frame, $^{90}$Sr-source, the DUT LGAD and the trigger LGAD all housed in a climate chamber.

A 2.5 GHz – 8 bit vertical resolution LeCroy WavePro 725Zi-A digital oscilloscope recorded the signals at a sampling rate of 40 GS/s, therefore with a time discretization of 25 ps. The digitization noise of the oscilloscope contributes a varying fraction to the overall noise depending on the vertical scale used, and special care had to be taken to minimize that contribution. The effect is minimized by the fact that a large vertical scale (which has larger scope noise) is used for large gain signals, at which point the jitter contribution is small (see Sec. 5 below).

### *$^{90}$Sr β-Telescope*

The laboratory setup with $^{90}$Sr β-source is shown in Figure 2. A frame aligning the source, the device under test (DUT) and the trigger counter is placed in a climate chamber. The board housing the DUT has a hole in correspondence to the active area of the DUT so that the electrons from the β-source are not stopped in the board material and can reach the trigger board. The dimension of this hole therefore defines the area of the sensor used in the testing. Because of the large radiation-induced leakage current, the irradiated LGAD were operated at -20 $^{o}$C and -30 $^{o}$C in a climate chamber that also provides shielding against electronic pick-up. The trigger and time reference is provided by a well-calibrated 50 µm thick un-irradiated CNM LGAD biased at 190 V, with a resolution of 29 ± 1.5 ps at -20 $^{o}$C and 35 ± 1.5 ps at +20 $^{o}$C determined from the measured time difference of two identical sensors.

The very low noise level of the trigger LGAD allowed triggering at about 7 times its noise RMS. Following a trigger, the impulses of both trigger and DUT were recorded, with a rate of a few Hertz. About 30% of the triggers consist of in-time signal coincidences between the DUT and the trigger LGAD, the majority of the rest was associated with noise, and a smaller fraction with multiple-scattered events. It should be noted that the coincidence rate DUT-Trigger was found to be stable for all fluences.

## 5 DATA ANALYSIS

The analysis follows the steps listed in [6]; additional details of the analysis can be found in [19] The digital oscilloscope records the full voltage waveform of both trigger and DUT in each event, so the complete event information is available for offline analysis. Average and normalized voltage pulses for three different fluences are shown in Figure 14 and Figure 21 below. The time of arrival of a particle is defined as the time at which the signal crosses a certain fraction of the maximum signal amplitude, this method, called constant fraction discriminator (CFD), is a very efficient tool to correct for the time walk effect. CFD circuits can be easily implemented in an ASIC. The main difference to previous measurements is the optimization of the CFD value for every bias voltage and fluence, a procedure that is necessary since both the pulse shape and the noise contributions change with fluence.

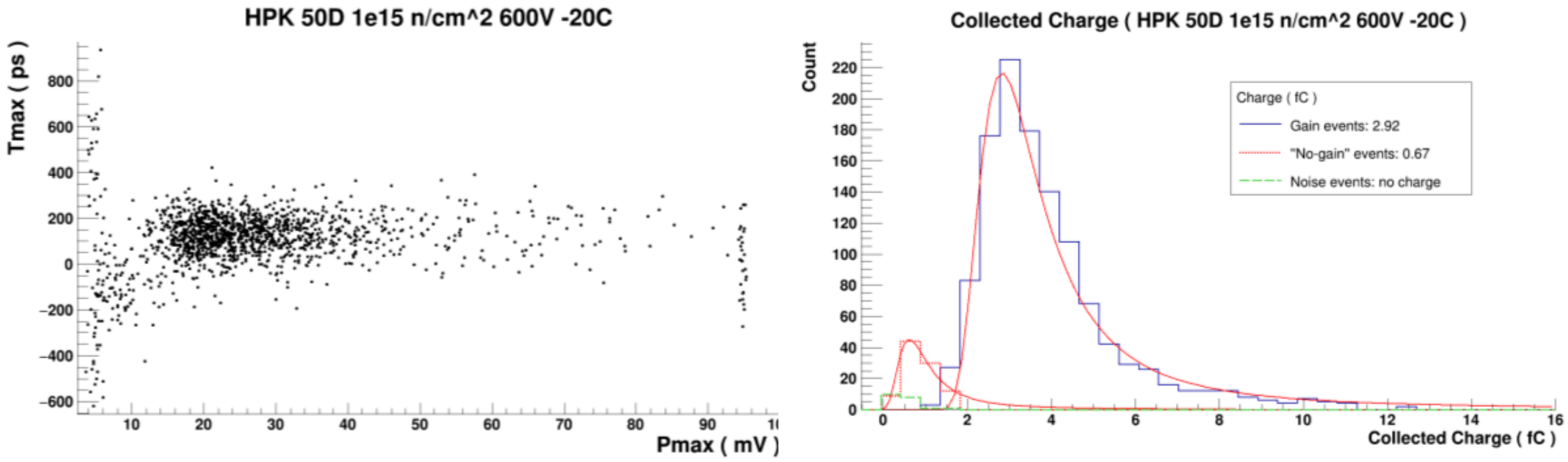

Figure 3 Scatter plot of the time of the pulse maximum, Tmax, vs. the maximum pulse height, Pmax, (left) for the LGAD exposed to a neutron fluence of 1e15 n/cm$^2$. The distribution of the integrated charge for the signals close in time to the trigger shows the form of a Landau distribution (right). The slightly out-of-time no-gain events appear at low pulse height.

The event selection is straightforward: for a valid trigger pulse, the signal amplitude, Pmax, of the DUT UFSD should not be saturated by either the scope or the read-out chain. To eliminate the contributions from non-gain events or noise, the time of the pulse maximum, Tmax, has to fall within a window of 1 ns centered on the trigger time. Figure 3, left, shows the distribution of the Tmax and Pmax, the maximum amplitude of a pulse, of HPK 50D. The distribution is dominated by large (gain) events, happening 100-200 ps after the trigger time, however a small fraction of small (no-gain) events are seen at early times, indicating that a fraction of the electrons are hitting the DUT edge, where there is no multiplication. The plot also shows that the fraction of saturated signals, due to the limited range of the oscilloscope, is negligible.

Integrating the voltage pulse and dividing it by the trans-impedance of the amplifier measures the collected charge for each signal. The distribution of collected charge is then fit to a Landau curve and the charge MPV is extracted, an example is shown in Figure 3, right.

It is important to note that even in thin sensors the fraction of the initial charges created by a MIP collected by the electronics decreases, due to trapping, with increasing fluence. This effect is shown in Figure 4, which shows the collected charge for a PIN sensor as a function of fluence when biased below 500V where no multiplication in the bulk is observed. The data was taken with a 60 micron thick PIN [21] fabricated by Fundazione Bruno Kessler (FBK). The trapping time parametrization used previously in the WF2 program [23] was updated to fit the data, and the expected fluence dependence of the collected charge for 60, 50 and 35 micron thick sensors calculated with WF2 are shown. As shown the effect of charge trapping is quite benign at high fluences [29], and consistent with limited trapping rates at high fluences observed in [30].

In the following part of this work, the gain is obtained by dividing the measured MPV value of the DUT by the collected charge of a standard silicon detector (PIN) of the same thickness and irradiated to the same neutron fluence, biased at 500V, i.e. without any internal multiplication, as shown in Fig. 4. In this definition, the LGAD gain is therefore the sum of the gain due to the multiplication in the gain layer and that in the bulk due to the applied bias voltage. The contribution of the uncertainty of the PIN charge to the error in the gain determination is estimated to be 5%, the size of the error bars in Fig. 4 (in addition to the common systematic scale error of 20% in the trans-impedance mentioned in Section 0).

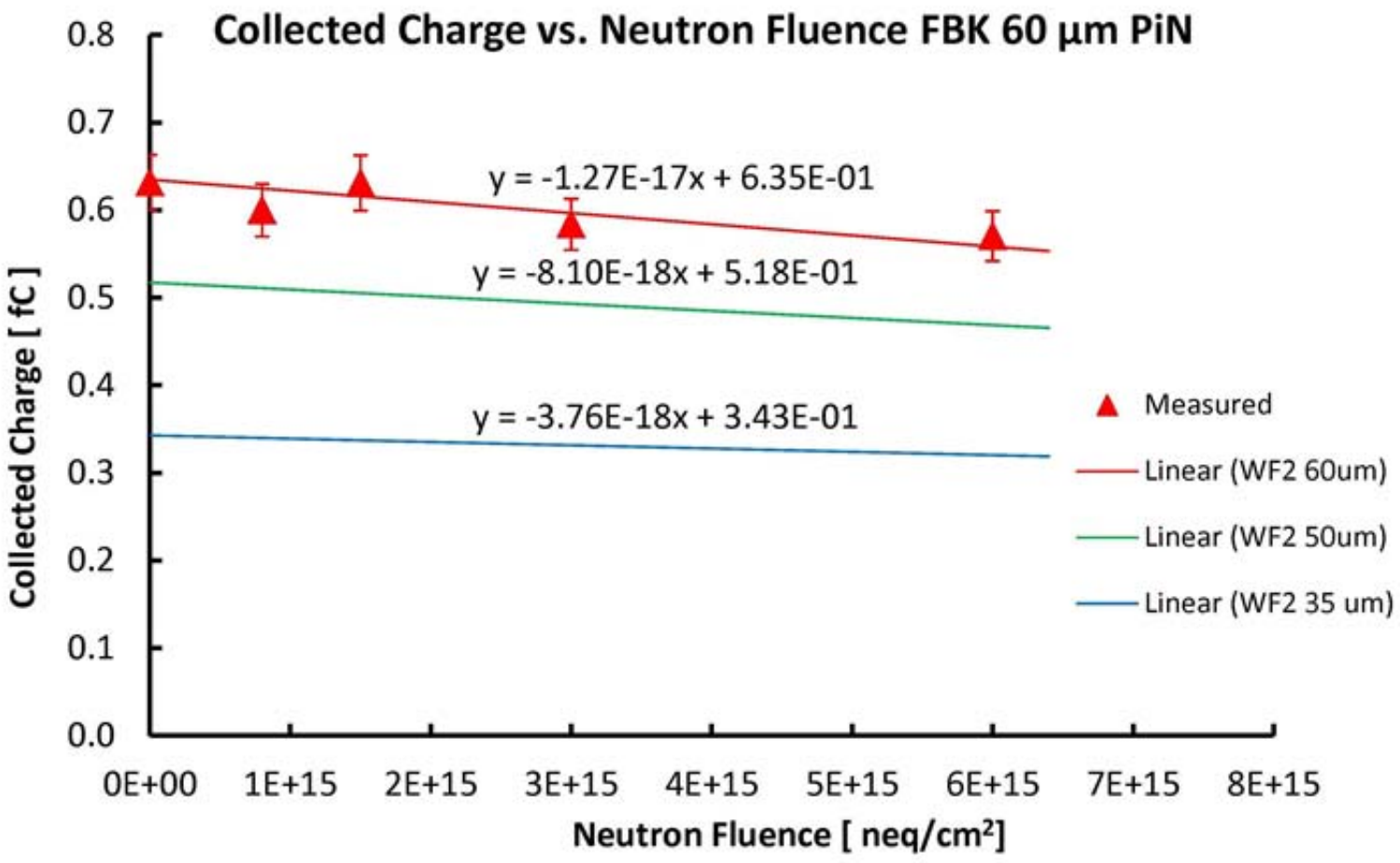

Figure 4 Charge collected from a MIP in a thin p-type PIN silicon detector as a function of neutron fluence. The data were taken with a 60 micron thick PIN [21] fabricated by Fundazione Bruno Kessler (FBK). The trapping time parametrization used previously in the WF2 program [23] was updated to fit the data, and the expected fluence dependent collected charge for 60, 50 and 35 micron thick sensors calculated with WF2 are shown.

Due to the oscilloscope digitization steps, the time of arrival at a specific CFD value was extracted from the recorded samples by means of a linear interpolation between the points above and below the requested value. By measuring the timing resolution as a function of the CFD setting, the value of CFD threshold which minimized the resolution was found. This analysis was repeated at each fluence resulting in an improved timing resolution.

The DUT time resolution ($\sigma_t$) is calculated from the RMS value ($\sigma_{\Delta t}$) of the Gaussian fit to the time difference Δt between the DUT and the trigger, both corrected for time walk with its proper CFD level:

$$\sigma_t = \sqrt{\sigma_{\Delta t}^2 - \sigma_{trig}^2}, \qquad (1)$$

with the resolution of the trigger being $\sigma_{trig}$ = 29 ± 1.5 ps at -20 °C and 27 ± 1.5 ps at -30 °C.

## 6 Results

### *Gain*

The UFSD gain as a function of bias voltage for different neutron fluences are shown in Figure 5 for operating temperatures of -20 °C and -30 °C. The bias required to reach a certain gain value increases with increasing fluence due to the acceptor removal mechanism [15]: as the gain layer becomes less doped and generates a weaker electric field, the external bias voltage needs to be increased to compensate for this loss. The highest bias voltage which can be reached before the sensor exhibits instabilities is called the "operating voltage $V_{op}$" and will be discussed in Session 0. For the LGAD 50D the gain at $V_{op}$ is below 20 at fluences beyond 6e14 n/cm$^2$ and below 10 beyond 3e15 n/cm$^2$.

The simulations program Weightfield2 [9] incorporates the parameterization of the measured acceptor removal mechanism [15] and predicts that when a bias value of about 600 – 650 V is reached, the electric field in the bulk becomes high enough to generate multiplication in the sensor bulk, moving the location of a part of the multiplication mechanism from the gain layer into the bulk.

For the high fluence points of 1e15 n/cm$^2$ and 6e15 n/cm$^2$ the bias voltage dependence of the UFSD gain is the same for T = -20 °C and T = -30 °C operation.

In the following, measured parameters will be presented as a function of both the bias voltage and the gain.

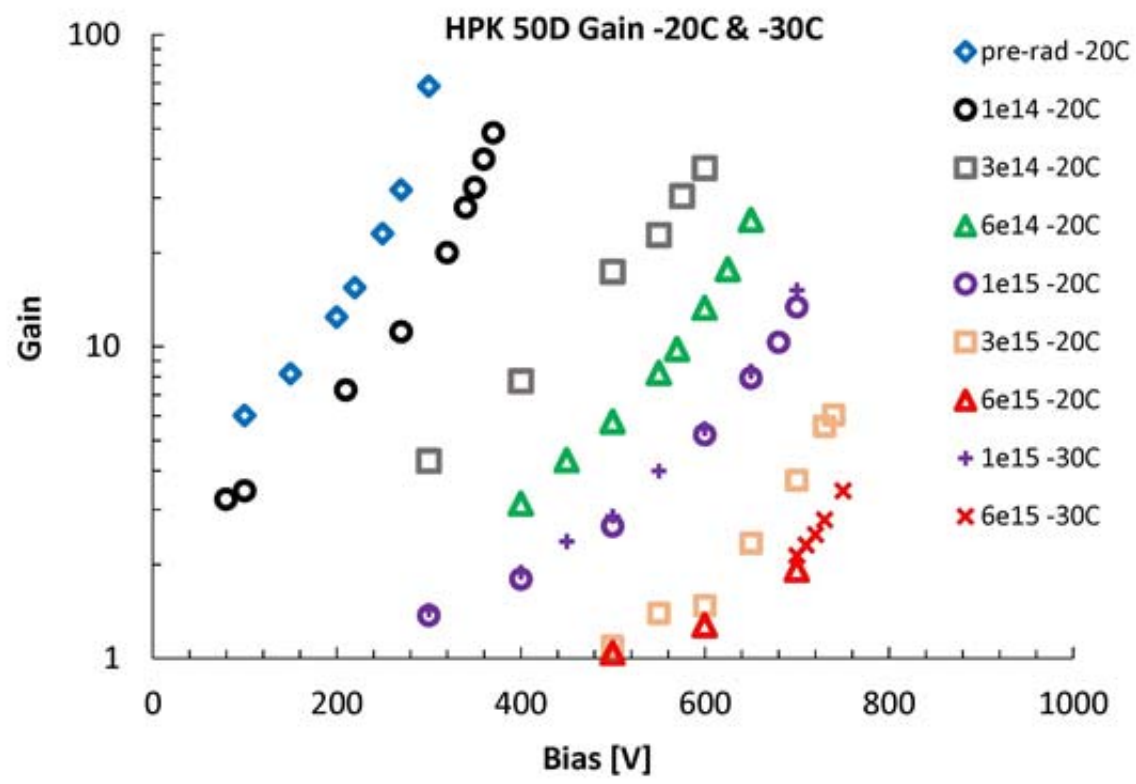

Figure 5 Gain as a function of bias of the HPK 50D LGAD irradiated to the indicated neutron fluences at -20 $^{o}$ C and -30 $^{o}$ C, showing the need for increasing the bias of irradiated sensors to reach adequate gain. The gain has a common systematic scale error of 20%.

*Leakage current*

Varing the temperature on several LGADs, we measured that the leakage current of LGADs depends exponentially on the temperature (about factor 2 increase for every 7 $^{o}$C degree increase); following the well-known behaviour of PIN sensors.

The dependence of the leakage current and gain on the bias voltage, at many fluences, is shown in Figure 6 where it is visible that at each fluence the leakage current exhibits large increases at both higher bias and higher gain. In order to investigate this behaviour further, two fluences (3e14 n/cm$^2$ and 6e15 n/cm$^2$) were studied in more detail. The leakage currents collected at the guard ring (IGR) and at the n-electrode (IBR) were measured separately in order to isolate the contribution of the gain mechanism, which is present only in IBR. Dividing IBR by the measured gain G, IBR/G, the value of current in absence of multiplication was extracted. For both fluences we observed that even once the effect of gain is removed, the current collected at the n-electrode, shown in Figure 7, exhibits (i) a strong bias voltage dependence and (ii) its value is larger than the estimate obtained with the usual ansatz of $\Delta I = \alpha(T)$*Vol*Fluence, where $\alpha(T)$ is the current damage constant adjusted for the operating temperature (alpha(-20 $^{o}$C) = 6.7e-19), and Vol the detector volume.

This effect has been reported in [12] and seems to be unrelated to the gain mechanism. Further studies are needed.

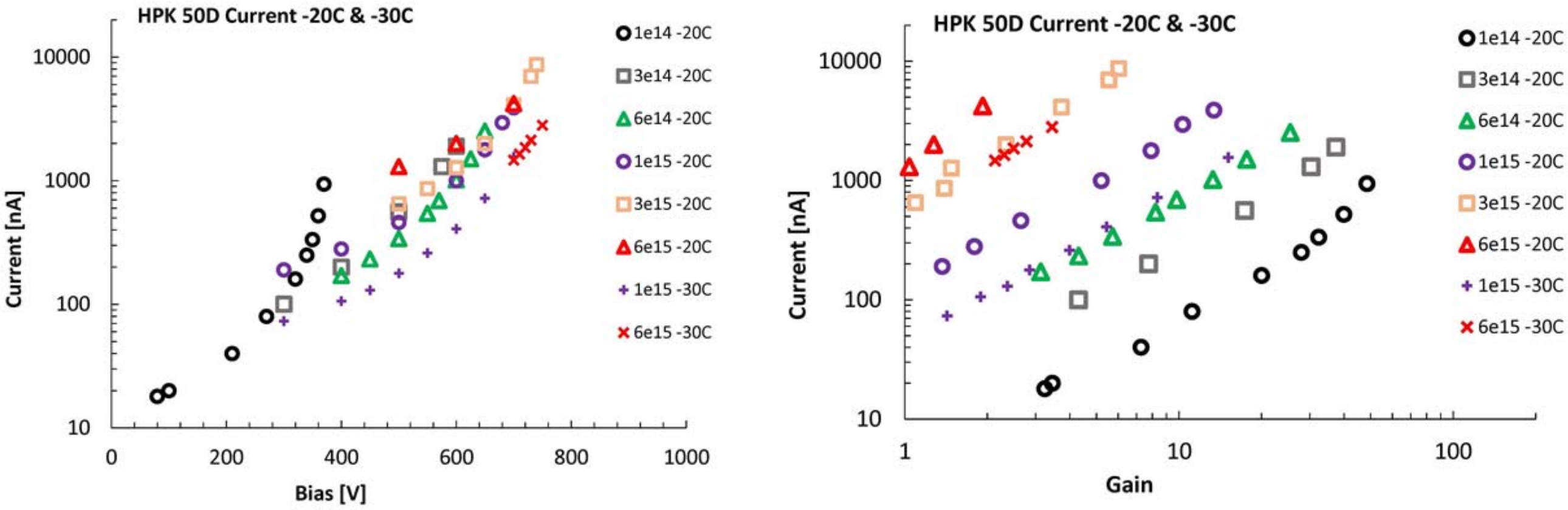

Figure 6 Leakage current at -20 $^{o}$C and -30 $^{o}$C as a function of bias (left) and gain (right) of the LGAD test structures irradiated to the neutron fluences and temperatures indicated.

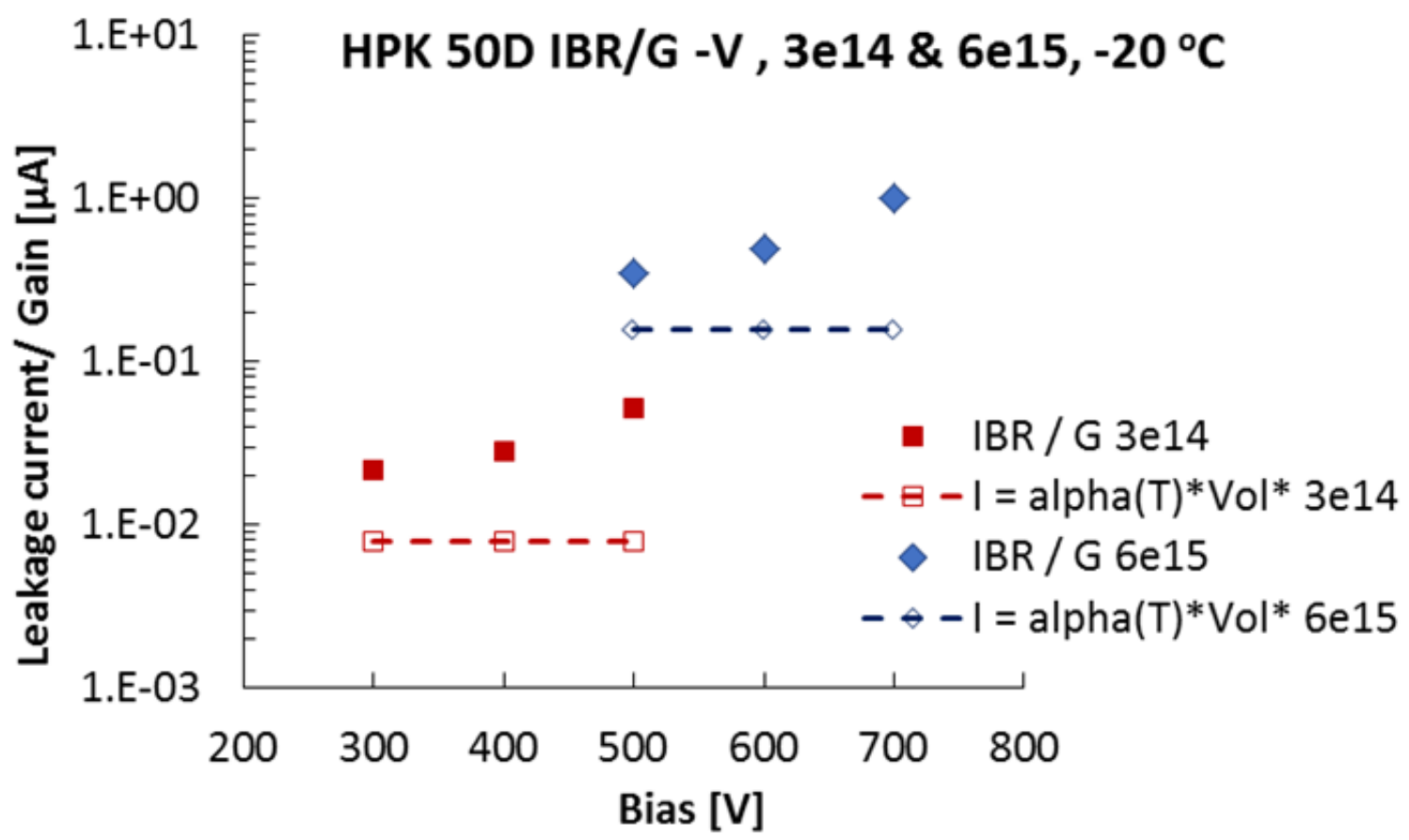

Figure 7 Bulk leakage current IBR divided by the gain G as a function of bias voltage for two fluences, 3e14 and 6e15 n/cm$^2$ ,at -20 $^{o}$C. Also shown is the result of the simple parameterization of the leakage current for PIN sensors without gain, with alpha(T) = 6.7e-19 at - -20 $^{o}$C.

*Time resolution: Gain dependence*

Following [1], the timing resolution $\sigma_t$ can be parameterized as

$$\sigma_t^2 = \sigma_{Jitter}^2 + \sigma_{LandauNoise}^2 + \sigma_{Distortion}^2 + \sigma_{Timewalk}^2, \qquad (2)$$

where the main terms contributing are (i) the jitter,

$$\sigma_{Jitter} = \frac{N}{dV/dt} \approx \frac{\mathrm{t_{rise}}}{\frac{S}{N}} \sim \mathrm{t_{rise}} \frac{\mathrm{N}}{\mathrm{G}} , \qquad (3)$$

which depends on the signal voltage amplitude S, the signal voltage slope dV/dt, the noise RMS N, the rise time $t_{rise}$ and the inverse of the gain G, and (ii) the "Landau noise", the fluctuations in the ionization profile. Signal distortion is minimized by the use of a "parallel plate" geometry for the sensors, while the use of the CFD method minimizes the time walk.

Figure 8 shows the MPV of the distribution of event-by-event jitter $\sigma_{Jitter}$=N/(dV/dt) at a CFD threshold of 20% as a function of the signal-to-noise ratio S/N for all fluences. In the log-log plot on the right side, the reduced slope for small S/N is an indication of the smaller rise time at large fluence (see Fig. 15).
A value for S/N > 20 is needed to lower the jitter below 20 ps.

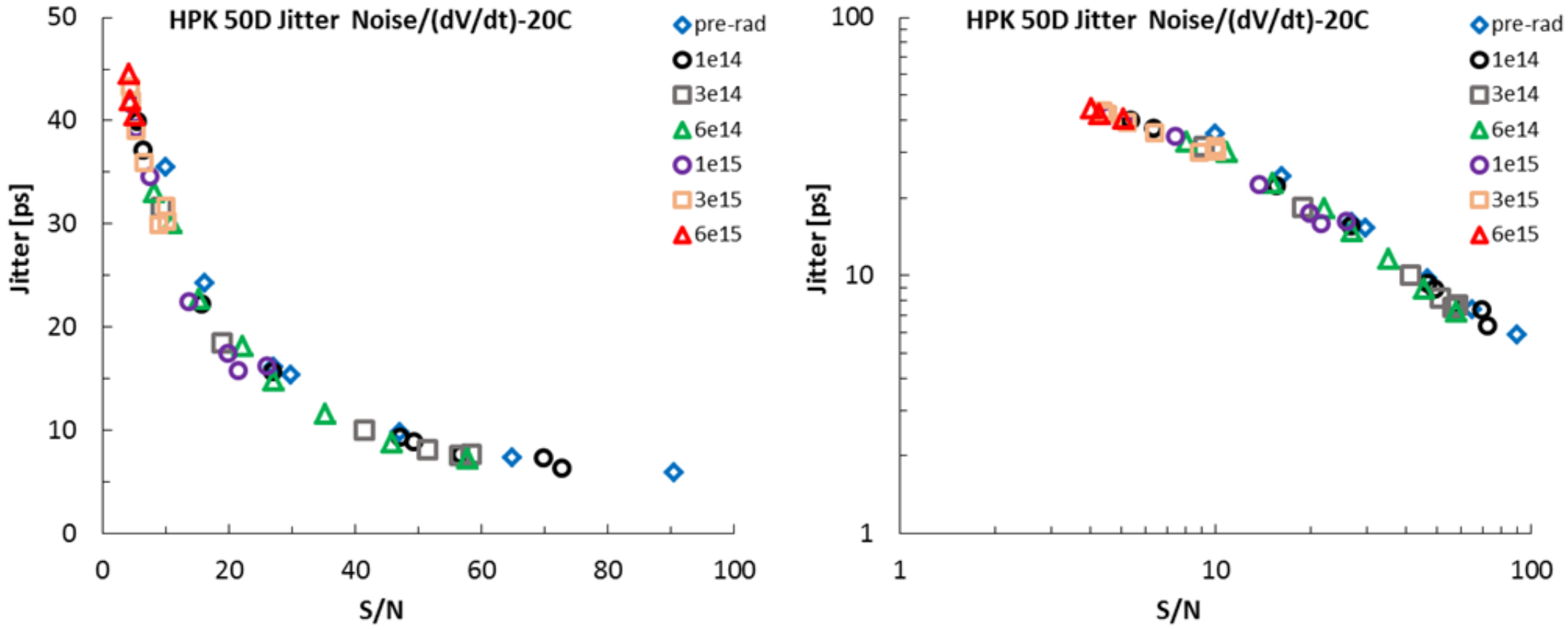

Figure 8 Jitter Noise/(dV/dt) as a function of the signal-to-noise ratio S/N

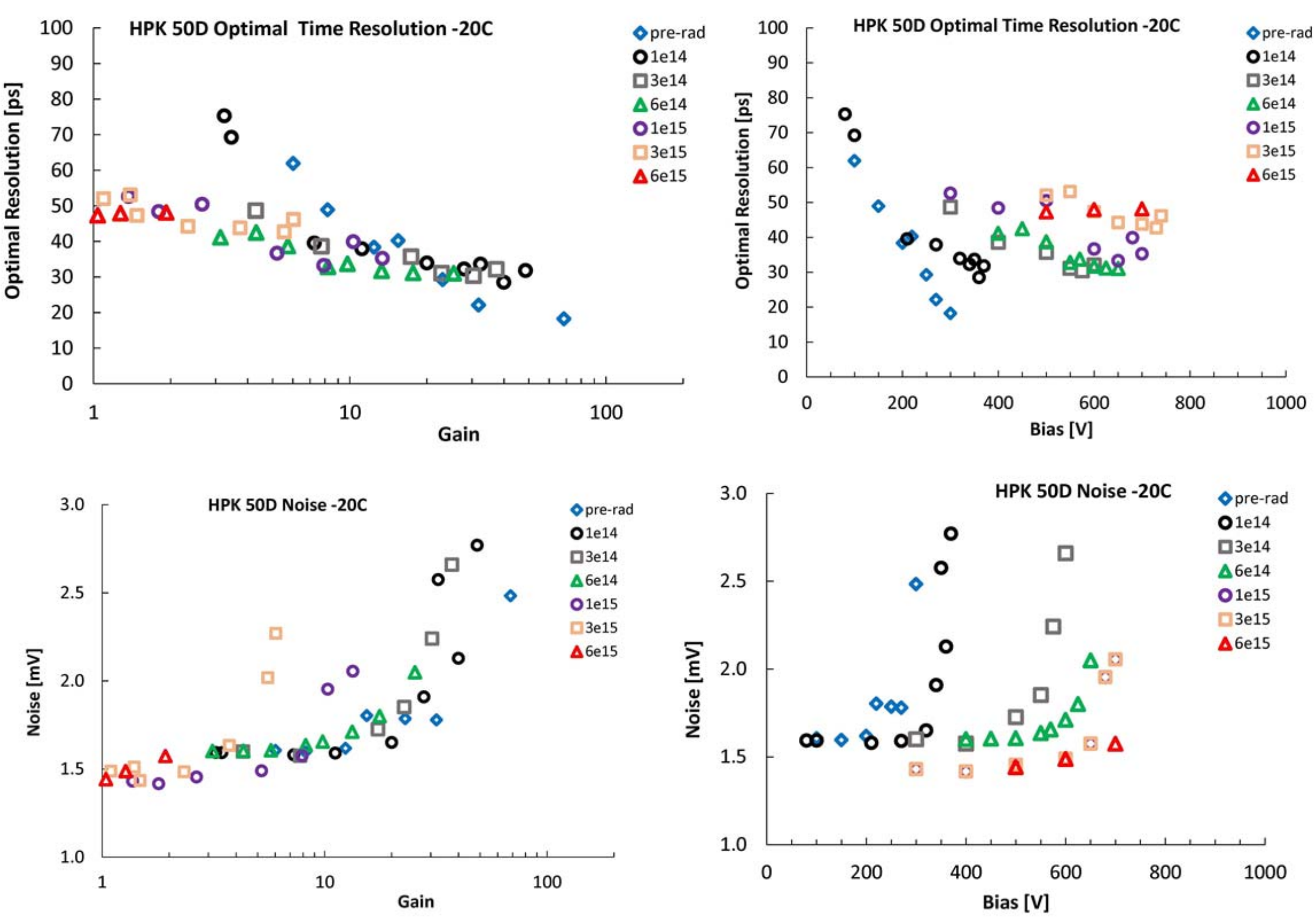

Figure 9 Time resolution evaluated at the optimized CFD fractions and the noise RMS vs. gain and bias for the different fluences at -20 $^{o}$C. The correlation between flattening of the resolution curve and the increase in noise is clearly visible.

In Figure 9 the time resolution using optimized CFD value are shown together with the noise both as a function of gain and as a function of bias voltage. The time resolution is initially falling rapidly with increasing gain, to reach almost a plateau for gain above 20. In the plot, it is possible to identify at low gain two different groups, with the first one formed by the lower-fluence data up to 1e14 n/cm$^2$ and the second by higher-fluence data. For a fixed value of gain, for example gain = 5, the lower-fluence group has a worse time resolution than the higher-fluence group. This fact, which might seem counter-intuitive, is due to the different type of multiplication mechanism in the two groups: as mentioned in Section 0, below a fluence of 1e15 n/cm$^2$ the multiplication occurs mainly in the gain layer, while for the higher fluence samples the multiplication happens everywhere in the bulk, allowing for the multiplication process to start earlier. The observed decrease of the rise time at higher fluence contributes to the improved time resolution (see below). The fact that for every fluence the time resolution flattens somewhat at the highest gain reflects the fact that for low fluences the resolution is controlled by the Landau noise, and for high fluences it is controlled by the jitter contribution, which is sensitive to the observed increase in noise.

### *Jitter and Landau Fluctuation*

Irradiation changes the gain and the operating voltages, which in turn influence the magnitude of both the Landau noise and the jitter of Eq (2).

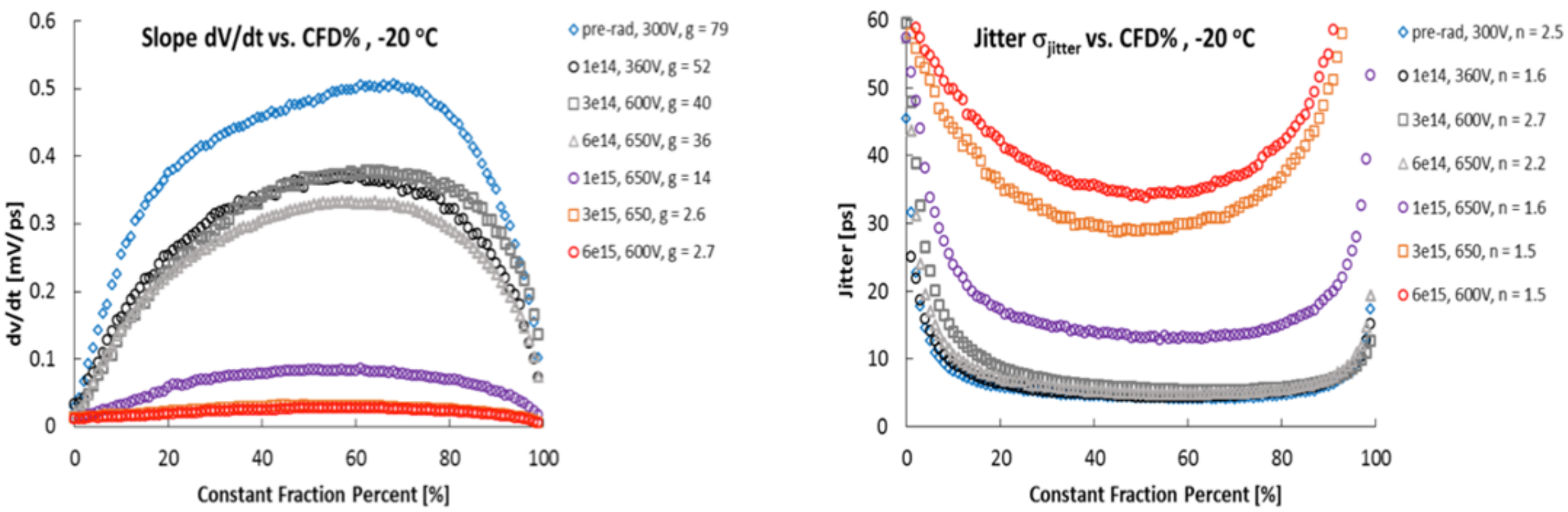

Figure 10: Measured slope (left, with g=gain) and jitter (right, with n=noise) as a function of CFD fraction for several fluences. For low fluences and high gain, the contribution of the jitter to the time resolution is small for CFD settings > 10%. For high fluences (low gain) the jitter increases and it is the dominant term in the time resolution.

The slope of the signal for different fluences, dV/dt, shown on the left side of Figure 10 is calculated at each point on the rising edge of the signal with a linear fit to the pulse shape using ±2 neighboring time bins. The slope is maximized for all fluences with a CFD setting > 50%. The jitter, extracted as the mean of the distribution of event-by-event values of $\sigma_{Jitter}$=N/(dV/dt) (using using Eq. (3), is shown on the right side in Figure 10. The jitter is lowest for the CFD set at ~ 50%. For those settings, the jitter depends largely on the gain which can be reached at the different fluences. Note that for fluences below 1e15 n/cm$^2$, the jitter is about the same even though the gains are quite different. The plot of the slope (Figure 10, left), reflects the difference in gain and rise time. For the jitter (Figure 10, right) the different noise RMS are contributing in addition. There is a limitation of the use of factorization as shown in Eq. (3): we find that the MPV of the distribution of event-by-event jitter values is quite different (in general lower) than the value of N/(dV/dt) calculated by dividing the mean of the noise distribution by the MPV of the dV/dt distribution.

Figure 11 shows the contribution to the time resolution due to non-uniform charge deposition (Landau noise) according to the simulation program WF2. Figure 11 (top) illustrates that this contribution is minimized by a low CFD setting for all sensor thicknesses. The bottom pane of Figure 11 shows instead the Landau noise contribution to the time resolution as a function of gain for different fluences: low CFD settings are better for all fluences and, remarkably, this contribution decreases with irradiation, as the gain layer doping is depleted. The circle indicates that for a gain of ~ 5 the contribution is roughly constant with irradiation.

The relative size of jitter and Landau noise determines the optimal CFD setting: for low fluences (high gain) the jitter is low and the larger Landau noise is minimized at low CFD settings. At high fluences (low gain) the Landau noise is low and the larger jitter is minimized by large CFD settings.

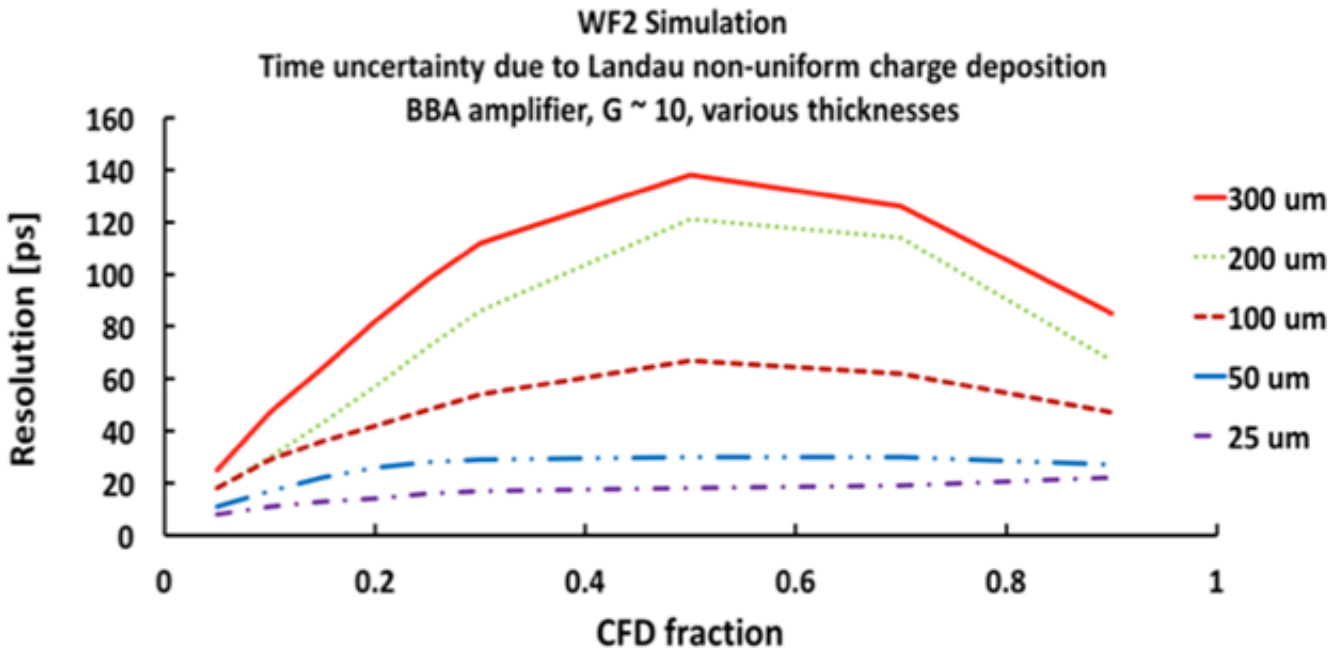

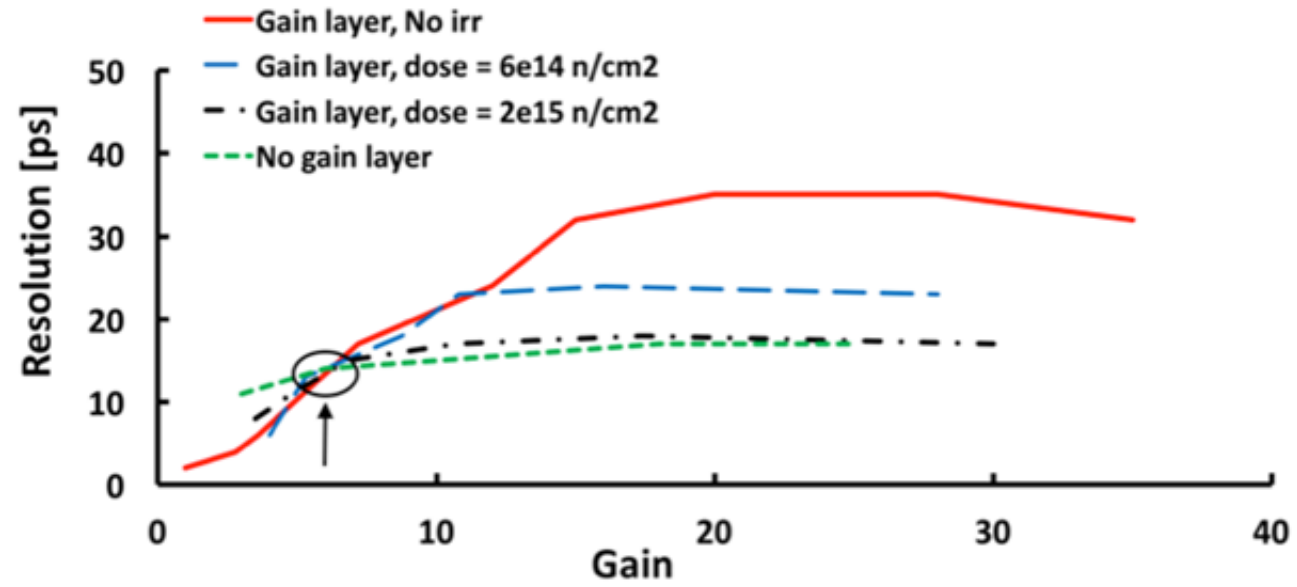

Figure 11 WF2 prediction of the contribution to the time resolution due to initial non-uniform ionization (Landau fluctuations): pre-irradiation as a function of CFD fraction for several thicknesses (top), and as a function of gain for CFD = 20% for several fluences, including no-gain sensors (bottom) for a 50 µm thick detector. The circle indicates that for gain values of ~ 5 the contribution is roughly constant with irradiation.

*Time resolution: CFD scans*

As Figure 14 below shows, the pulse shape changes as a function of fluence. This fact, combined with a fluence-dependent Jitter and Landau noise term (Section 0), suggests that a fixed CFD value will not minimize the time resolution for all fluences (and even different gains). To investigate this possibility, Figure 12 shows the values of the time resolution as a function of the CFD settings for different neutron fluences, demonstrating that the value of CFD that minimizes the time resolution is not a constant.

More details on this effect are shown in Figure 13 where each plot shows CFD scans for several bias voltages for a given fluence and temperature settings. A comparison between the CFD scans at -20 $^{o}$C and -30 $^{o}$C for the 6e15 n/cm$^2$ data shows a 10% improvement of the timing resolution with a temperature decrease of 10 $^{o}$C, while no such improvement is seen for the 1e15 n/cm$^2$ data.

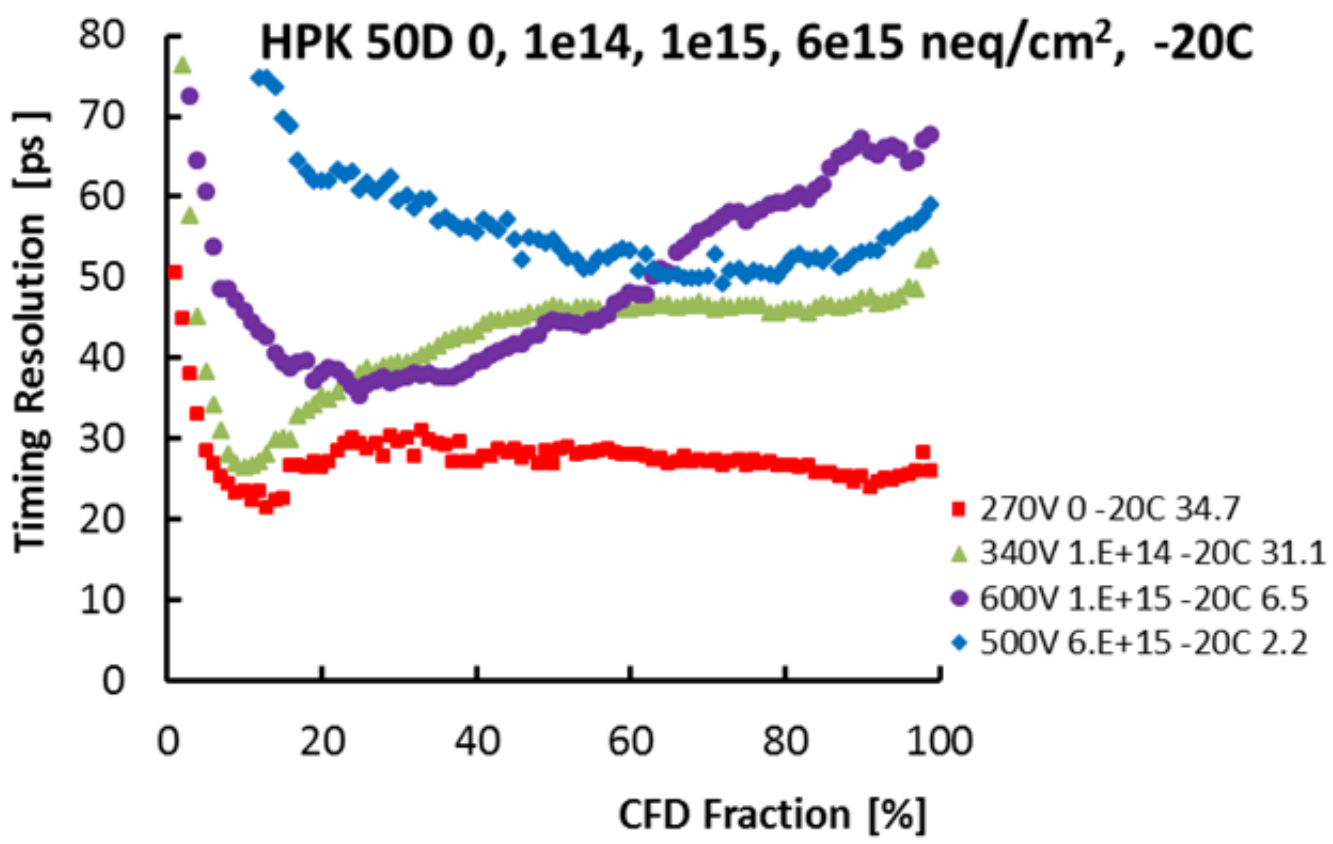

Figure 12 Time resolution for selected fluences as a function of CFD threshold at VHR bias voltages defined below. The legends here and in the following are bias voltage, fluence, temperature, gain.

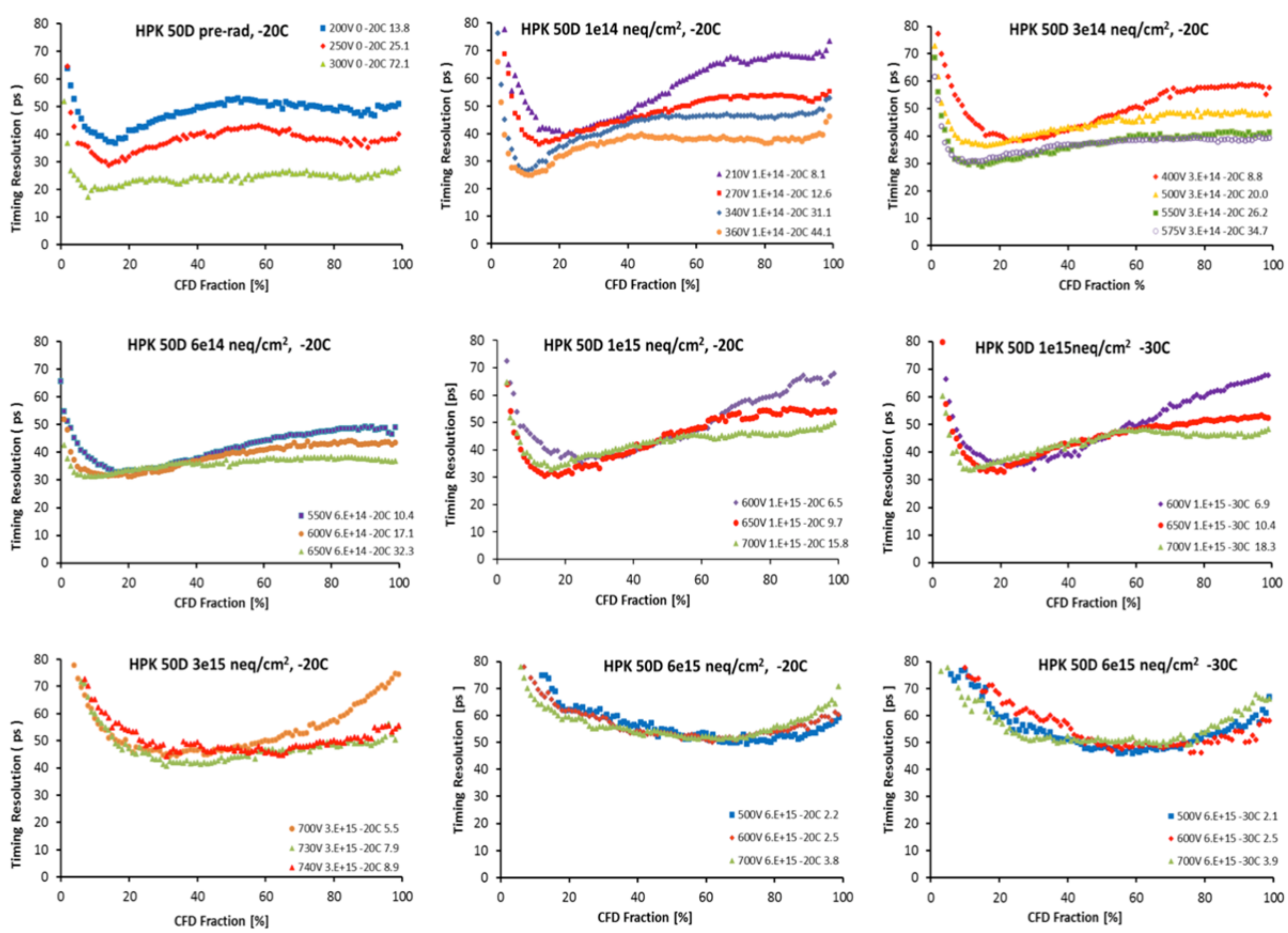

Figure 13 CFD threshold scans for several bias voltages for each of the seven neutron fluences. The data were taken at -20° C for all fluences and two fluence points (3e15 & 6e15 n/cm$^2$) were also measured at -30° C.

*Rise Time*

As shown above, the jitter depends on the gain and the rise time (c.f. Eq. (3)). The normalized average pulse for different fluences, Figure 14, shows that the pulse shape changes when the gain in the multiplication layer is reduced and the trapping is severe. The pre-radiation pulse is dominated by a "plateau current" due to the drift of the holes generated in the multiplication process, while at 6e15 n/cm$^2$ (i) the initial electrons and the gain electrons make the pulse front steeper and (ii) trapping reduces the pulse tail. Figure 15 shows the 10% - 90% rise time as a function of gain for the different fluences. Shorter rise times are observed at small gain, when the gain layer makes only a small contribution, and they increase when the gain is increased. For larger gain (at low fluences) the rise time becomes as long as the electron drift time [1]. As shown in Section 7, the fluence dependence of the rise time is well described by the WF2 simulation.

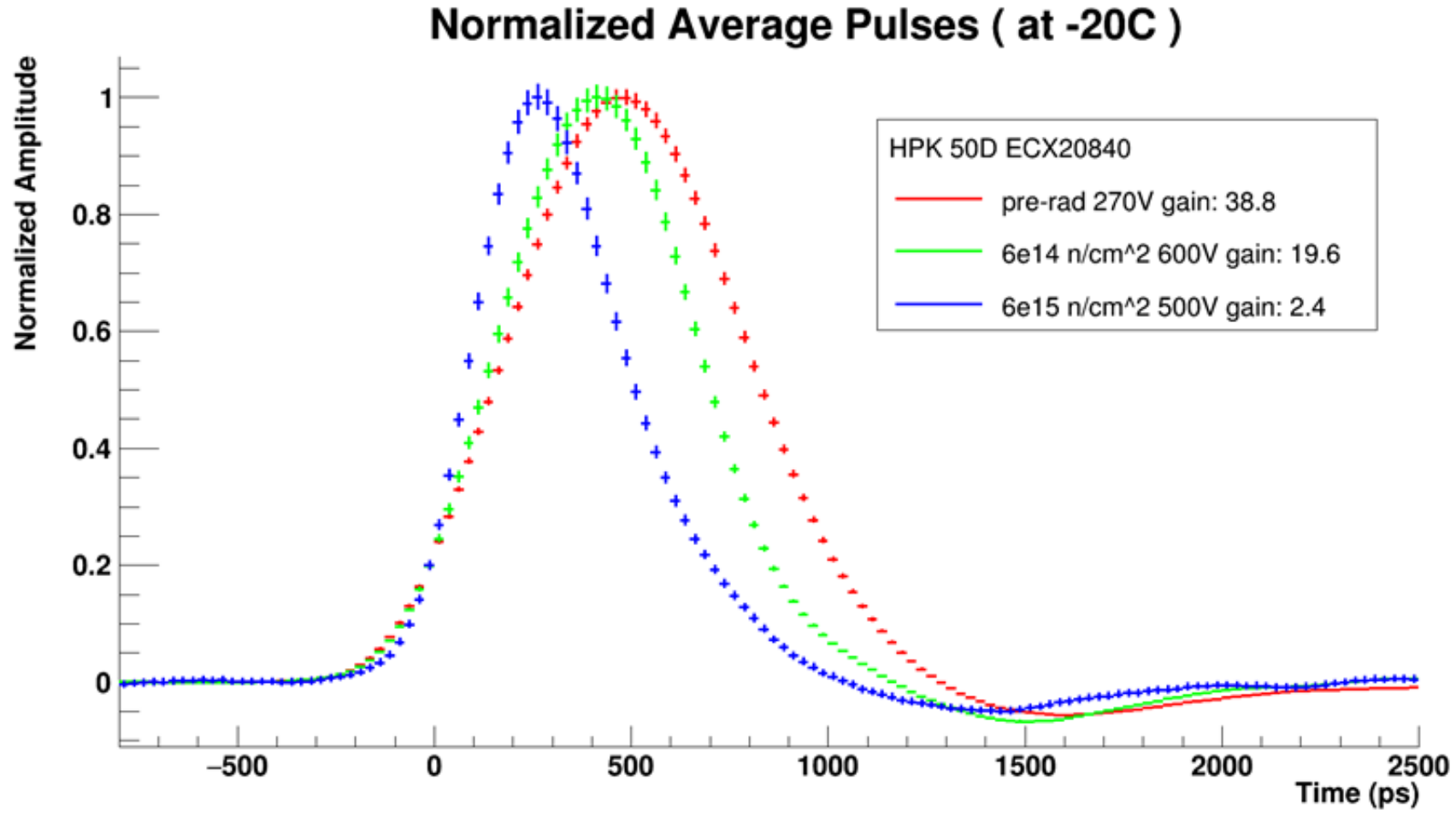

Figure 14 Averaged pulse shapes at three bias voltages pre-irradiation, and after fluences of 6e14 and 6e15 n/cm$^2$.

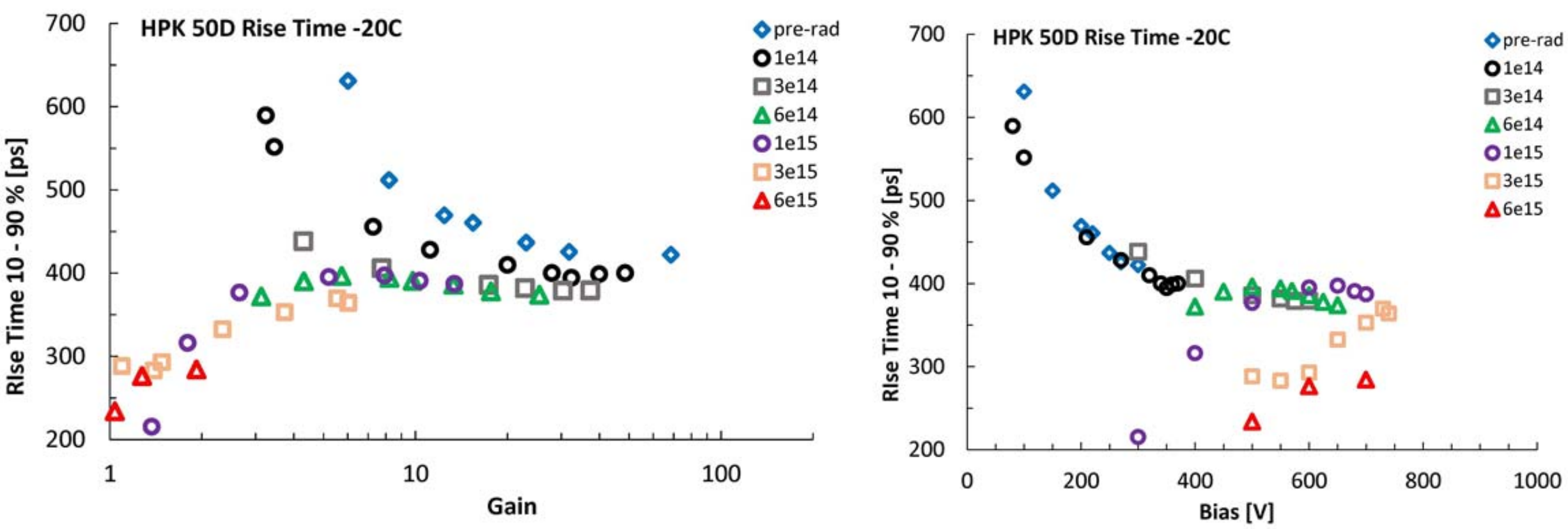

Figure 15 Rise time 10% to 90% vs. gain (left) and bias voltage (right) for the different fluences and temperatures. Note (i) the short rise time at small gain for irradiated sensors, which increases when the gain is increased and (ii) a convergence of the rise time at larger gain, expected from the saturation of the drift velocity with the bias voltage.

*Bias Reach – "Headroom"*

In order to gauge the performance of the LGAD during operations at the HL-LHC we consider the following two scenarios for biasing the sensors. The first one is to bias the sensors at the operating voltage $V_{op}$ , defined as the voltage that gives the best timing resolution for that fluence. This bias is lower than the break-down voltage, where the current becomes excessive and the resolution deteriorates. In order to find the sensitivity of the sensor performance to the accuracy of the setting of the bias voltage, we investigate in addition a second setting where the sensors are kept at a lower bias, called the "headroom voltage" $V_{HR}$. The difference between $V_{op}$ and $V_{HR}$ is simply called the headroom, and the relative headroom is $(V_{op} - V_{HR})/ V_{op}$. The values of $V_{HR}$ for each fluence can be derived from Figure 9 where at a bias below the operating voltage $V_{op}$ the resolution levels out and the noise increases. For example for a fluence of 6e14 between $V_{HR}$ = 600 V and $V_{op}$ = 650V the resolution is almost constant while the noise increases.

In Figure 16 and Table 1 the time resolution and the corresponding optimal CFD fraction are shown for both bias scenarios. While for un-irradiated sensors both bias settings are below 300 V, they increase for fluences greater than 6e14 $n/cm^2$ to 500 V or more (Figure 16, left). The relative headroom varies from 8% at low fluences to 18% at higher fluences. This is important since the optimal operating voltage is well known at the beginning of the data taking, while later on in the detector operation, after large fluences, there will be a certain amount of uncertainty on how to bias the sensors. The optimal CFD setting is constant at about 10-15% up to 1e15 $n/cm^2$, with a sharp increase above that number (Figure 16, right). As shown in Figure 17, the time resolution increases with fluence from 20 ps pre-radiation to 50 ps after the highest fluence with small (≈ 2 – 4 ps) increase when changing from $V_{op}$ to $V_{HR}$. If a fixed CFD threshold is required, according to Figure 17, a value of 30% is a good compromise.

Table 1. Fluence dependence of the operation bias $V_{op}$, and headroom voltage $V_{HR}$, and of the corresponding values for MPV of the collected charge, gain, timing resolution, the CFD fraction (CFD %) and headroom.

| Temp. [$^{o}$C] | Fluence [$n/cm^2$] | Vop [V] | Charge [e-] | Gain | Resol. at Vop [ps] | CFD% at Vop | VHR [V] | Charge [e-] | Gain | Resol. at VHR [ps] | CFD% atVHR | Headroom [V] | [%] |
|---|---|---|---|---|---|---|---|---|---|---|---|---|---|
| -20 | 0 | 300 | 2.30E+05 | 72.1 | 18.4 | 8 | 270 | 1.11E+05 | 34.7 | 21.0 | 13 | 30 | 10% |
| -20 | 1E+14 | 360 | 1.34E+05 | 43.0 | 27.6 | 10 | 320 | 6.90E+04 | 22.1 | 33.8 | 15 | 40 | 11% |
| -20 | 3E+14 | 600 | 1.26E+05 | 41.2 | 31.1 | 10 | 550 | 7.82E+04 | 25.5 | 30.8 | 15 | 50 | 8% |
| -20 | 6E+14 | 650 | 9.38E+04 | 31.9 | 30.9 | 10 | 600 | 4.97E+04 | 16.9 | 31.2 | 21 | 50 | 8% |
| -20 | 1E+15 | 700 | 4.43E+04 | 15.7 | 34.6 | 17 | 600 | 1.81E+04 | 6.4 | 36.0 | 25 | 100 | 14% |
| -20 | 3E+15 | 730 | 1.83E+04 | 8.3 | 42.4 | 46 | 600 | 5.35E+03 | 2.4 | 48.0 | 54 | 130 | 18% |
| -20 | 6E+15 | 600 | 4.36E+03 | 2.6 | 47.4 | 62 | 500 | 3.78E+03 | 2.2 | 45.4 | 66 | 100 | 17% |
| -30 | 1E+15 | 700 | 5.12E+04 | 18.2 | 36.3 | 17 | 600 | 1.94E+04 | 6.9 | 34.5 | 29 | 100 | 14% |
| -30 | 6E+15 | 600 | 4.46E+03 | 2.6 | 45.3 | 88 | 500 | 3.65E+03 | 2.2 | 42.2 | 58 | 100 | 17% |

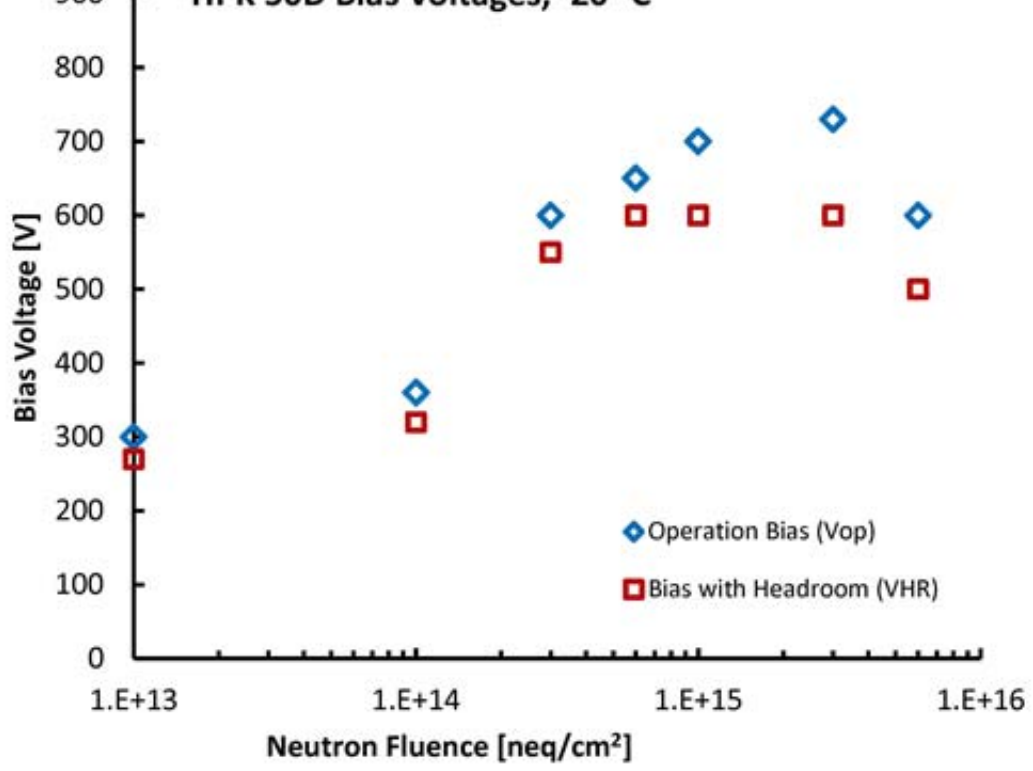

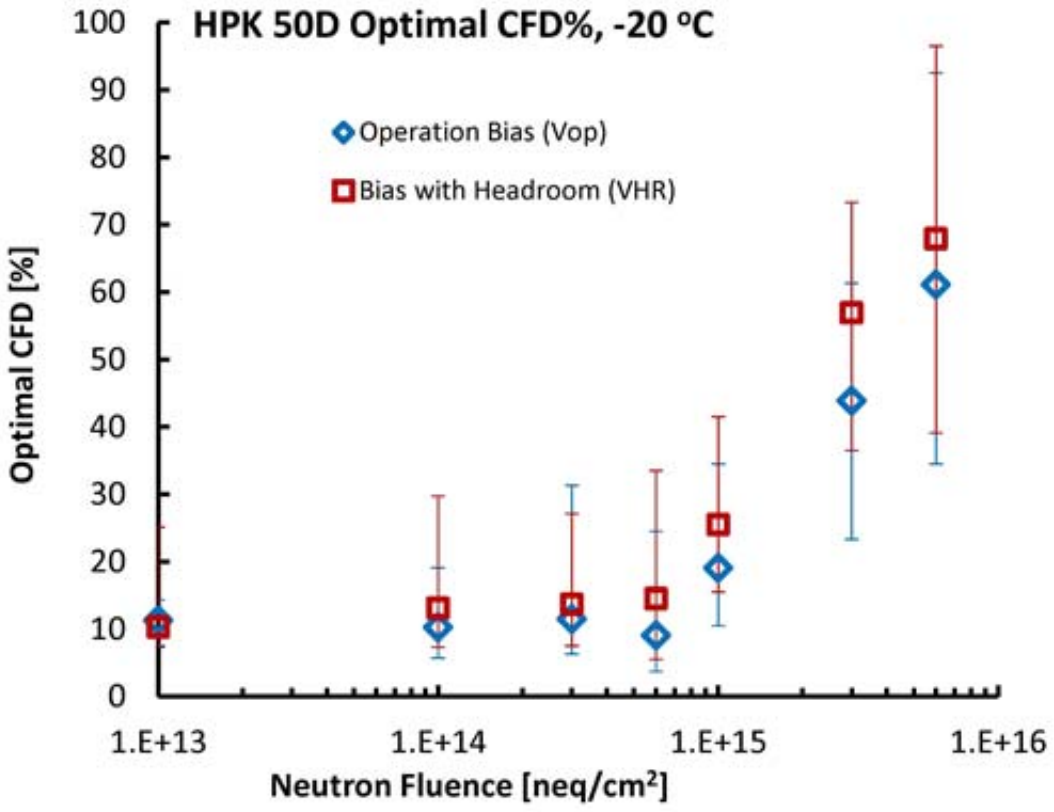

Figure 16 Fluence dependence of the operating voltage $V_{op}$ and head room voltage $V_{HR}$ (left), and the CFD fractions for best timing resolution at those two bias voltages (right), where the error bars indicate the range of the CFD threshold which yields a resolution within 10% of the optimal value. (For ease of presentation, the pre-irradiation point is shown at a fluence of 1e13).

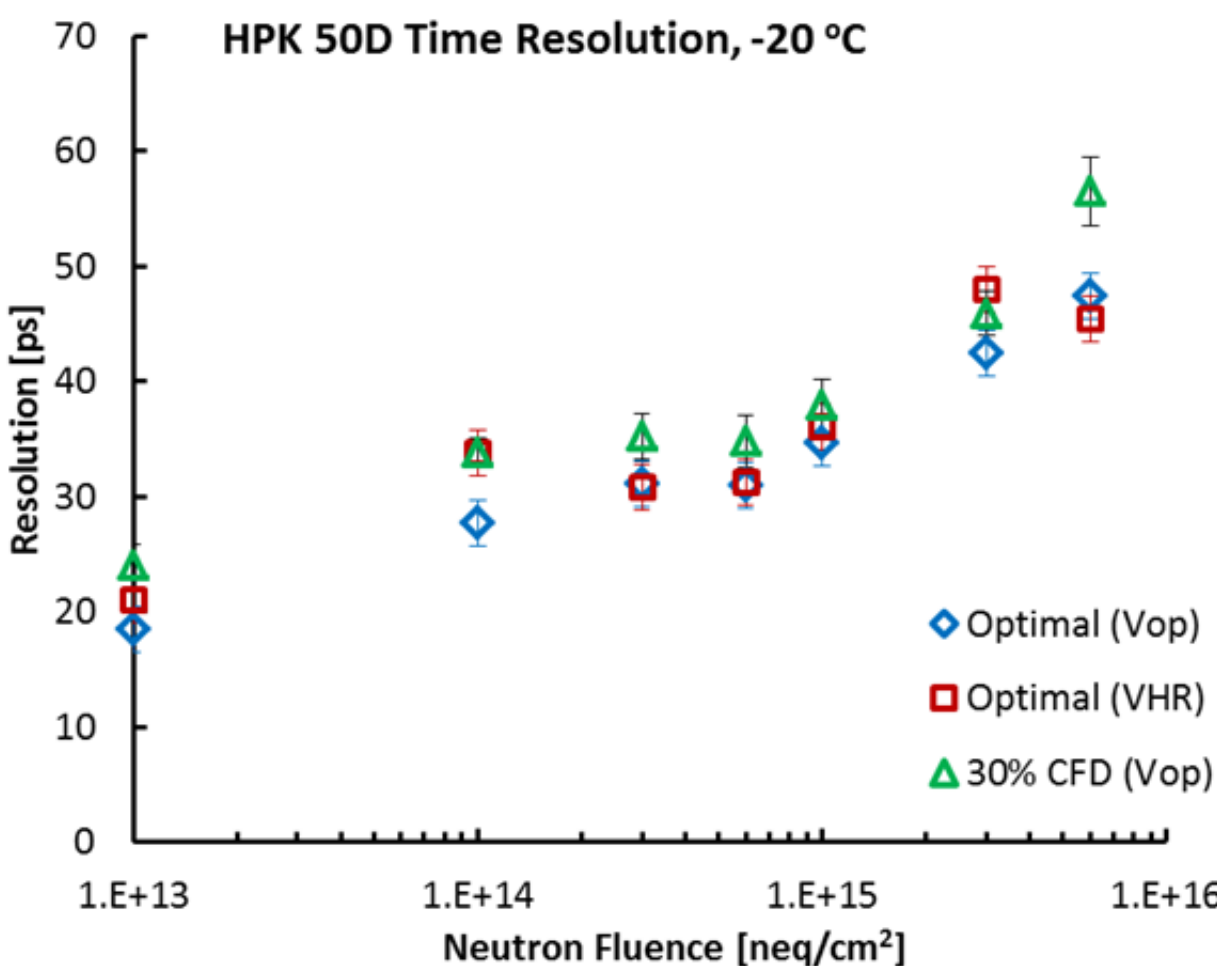

Figure 17 Fluence dependence of the time resolution at the two bias voltages Vop and VHR using the optimized CFD fractions and for constant CFD fraction of 30% for Vop. (For ease of presentation, the pre-irradiation point is shown at a fluence of 1e13).

Comparison with the results of [22] indicates that the superior bias reach of the HPK LGAD leads to improved time resolutions at comparable fluences and to a large extension of the useful fluence range.

*Doping Profile*

Capacitance-voltage (C-V) scans are used to investigate the radiation-induced changes in the doping profile of detectors.

Pre-irradiation scans are done at the customary frequency of 10 kHz at room temperature, and we find no frequency and temperature dependence by lowering the frequency down to 150 Hz and the temperature down to -30 $^{o}$C. For irradiated sensors, which require operation at lower temperature because of the increased leakage current, the frequency used has to be adjusted as shown in [24], and for operation at -20 $^{o}$C the value of 200 Hz was selected. The data with fluence of 6e15 n/cm$^2$ were taken at 150 Hz and -30 $^{o}$C.

The 1/C$^2$ scans for several fluences shown in Figure 18 indicate the two principal radiation effects for the 50D HPK sensors: (i) the voltage lag which signifies the depletion of the gain layer (also known as the "foot") decreases from 36 V pre-irradiation to 22 V after a fluence of 6e14 n/cm$^2$, to 19 V after a fluence of 1e15 n/cm$^2$, and to 3V after 6e15 n/cm$^2$, and (ii) the slope of the depletion curve of the bulk following the foot decreases significantly at increasing bias voltage, signaling an increased bulk resistivity.

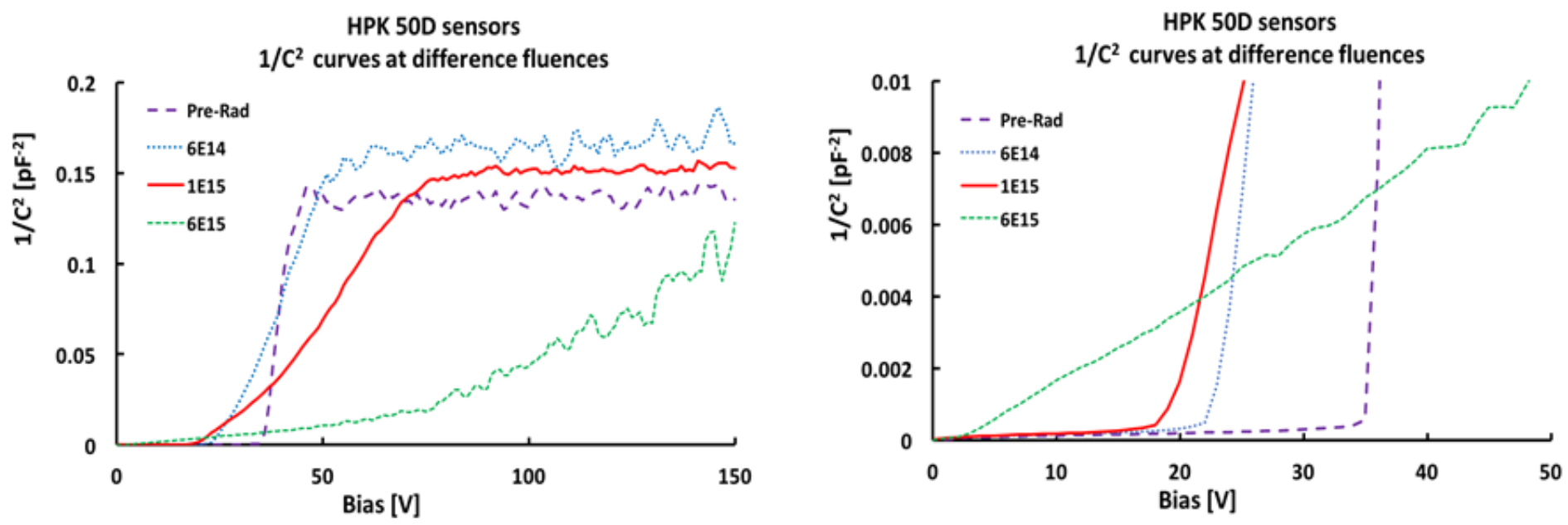

Figure 18 Bias scans of 1/C$^2$ for four fluences (zoomed on the right plot). The bias voltages lag ("foot") and the slope of the curves change with fluence, indicating both acceptor removal in the multiplication layer and acceptor creation in the bulk.

In order to extract information about the doping profile we take a phenomelogical approach and analyze the data assuming a uniformly doped substrate and a sharp semiconductor junction, which does not hold strictly in heavily irradiated sensors where trapped carriers affect the Space Charge Region (SCR). The small thickness of the sensors should mitigate these effects and also permit the use of the parallel plate capacitor approximation. Under these assumptions, below depletion the bias voltage $V_b$ and the depleted depth x are related to the doping concentration N by the following equation:

$$V_b = \frac{qN}{2\varepsilon} x^2. \tag{4}$$

Since the measured capacitance C depends on the depleted depth and the area A:

$$C = \varepsilon \frac{A}{x} , \tag{5}$$

the derivative of $1/C^2$ with respect to $V_b$ is proportional to 1/N:

$$N = \frac{2}{q\varepsilon A^2} \frac{1}{\frac{d\frac{1}{C^2}}{dV}} . \tag{6}$$

A C-V scan results in a N-V scan, and using eq. (5), in a N-x scan. For constant doping concentration the $1/C^2$ curve is linear as a function of the bias V, as indicated by eq. (6).
Using eq. (6), the doping profile for three fluences were extracted from the C-V curves, and they are shown in Figure 19. The two effects mentioned above are evident: the decrease of the doping concentration in the gain layer (driven by the acceptor removal mechanism), and the increase of the bulk doping (driven by the creation of acceptor-like defects).

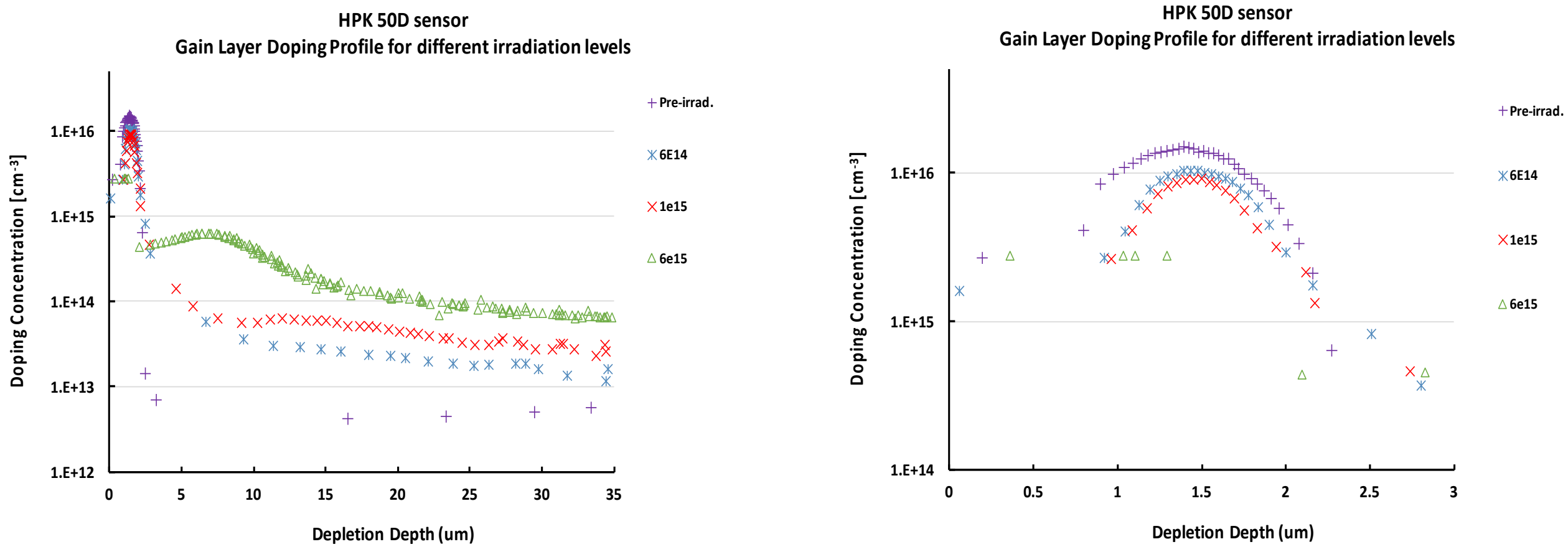

Figure 19 Extracted doping profile for four fluences. As explained in the text, the curves are obtained starting from the measured $1/C^2$ vs. bias plots.

In Figure 20 the observed evolutions of the doping concentration of the bulk and the multiplication layer are compared with the prediction of the model presented in [1]. For details of the comparison between measured points and model prediction as a function of the sensor depth, see [25]. As expected, the acceptor density increases linearly in the bulk once the density of acceptor-like defects exceeds that of the bulk, while in the multiplication layer it decreases exponentially.
Excellent agreements are evident for the decrease of the gain layer doping, while in the bulk, data of Figure 19 suggest a higher doping profile near the junction, reaching the value predicted by the model at a depth of 25 - 30 micron.

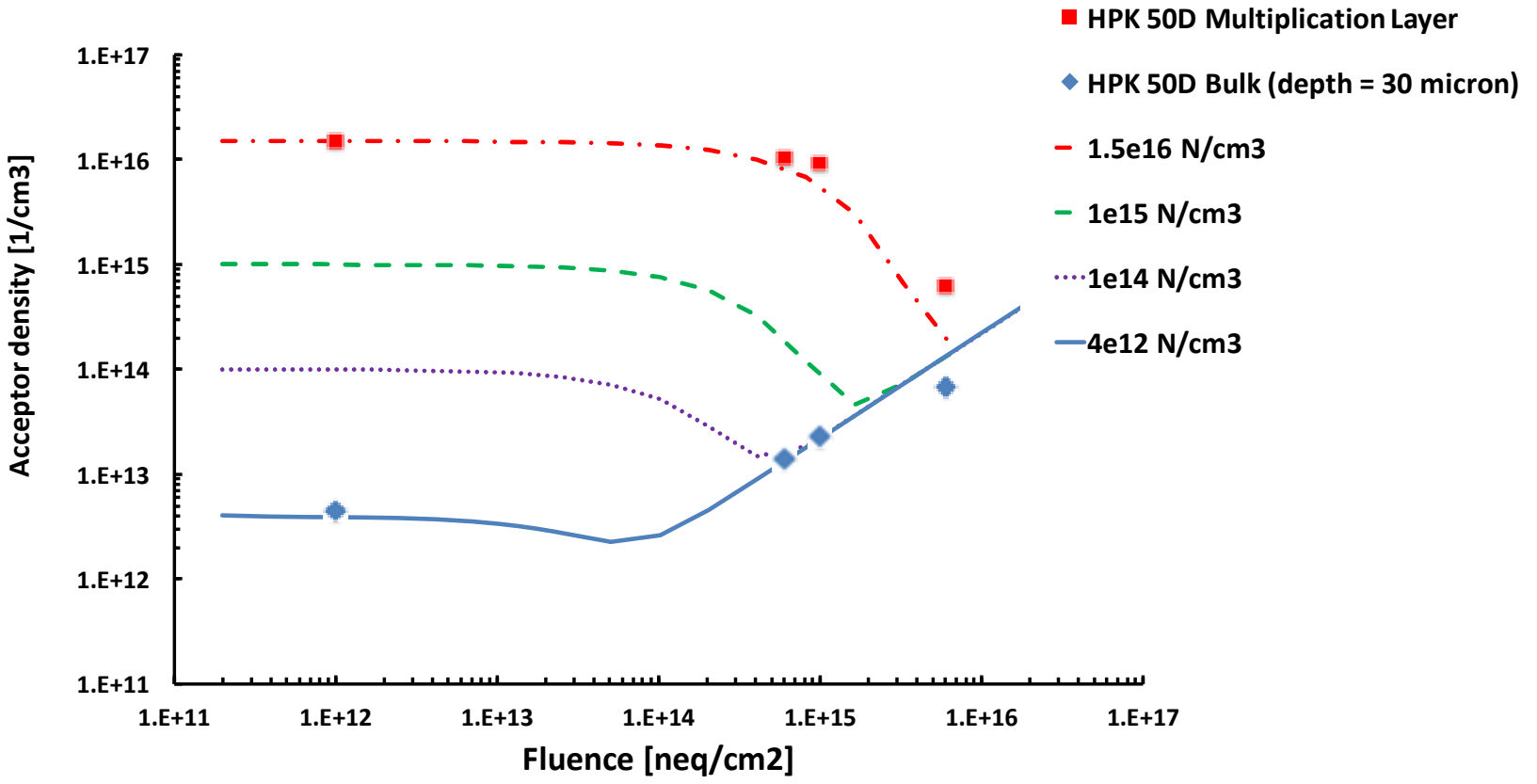

Figure 20 Evolution of the acceptor density in the bulk (blue) and in the multiplication layer (red) as a function of fluence, measured on HPK 50D sensors and predicted from [1].

## 7 Comparison with WF2 Simulation

In this section, the capability of the simulation program WeightField2 (WF2) to reproduce the key features shown above is explored. WF2 includes the effects of radiation on charge trapping, as explained in [26], and acceptor removal and acceptor creation via deep traps, as explained in Chapter 5 of [1]. The carriers' drift velocities and the impact ionization mechanism, and their dependences on the electric field and temperature, are included in WF2 using parameterizations taken from the Synopsis Sentaurus manual [27] and from [28].

Figure 21 shows the comparison between data and WF2 of the pulse shape at representative fluences in absolute values (left side) and normalized (right side). Both the decrease of pulse height, the rise time and the reduction of the tail with fluence are well modelled.

Comparison between data and WF2 predictions of the signal slope (dV/dt) vs. gain, the gain vs. bias and the rise time vs. gain are shown in Figure 22. In these variables there is a good agreement between data and WF2

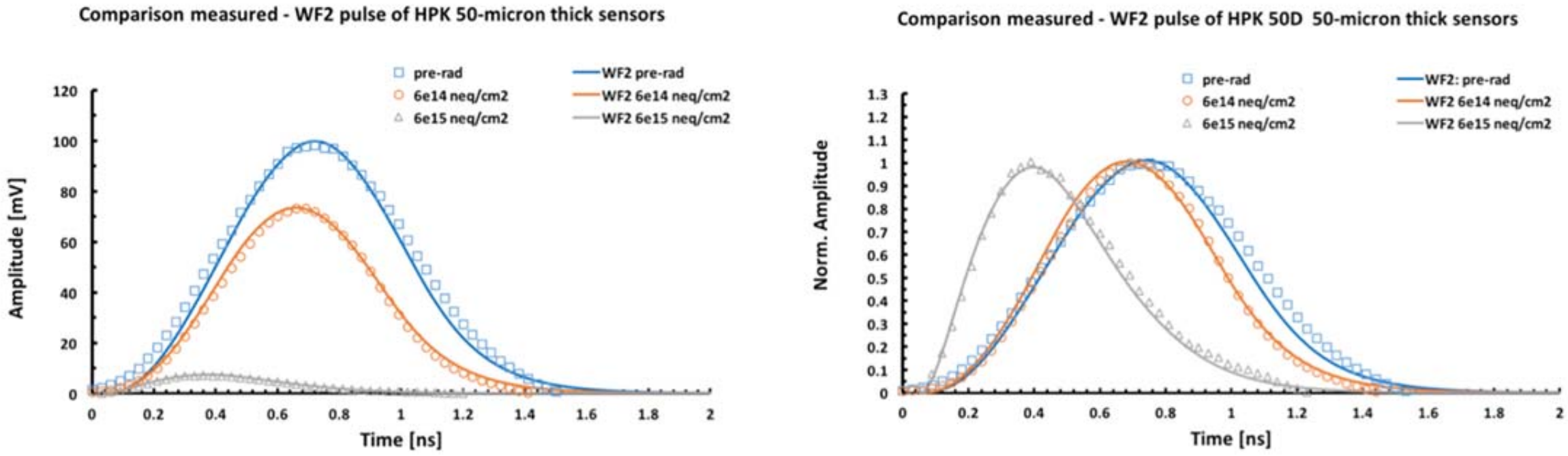

Figure 21 Data and WF2 prediction of the average pulse shapes for three fluences showing the reduction in gain (left); normalised pulse shapes, showing the decrease in rise time and width of the pulse (right).

simulation both in absolute values and in the functional trends as a function of gain and bias.

These comparisons are therefore important as they confirm our capability of modelling the dynamic mechanisms of gain including the interplay of the multiplication mechanism in the gain layer and in the bulk, and the effects of radiation damage in UFSD sensors.

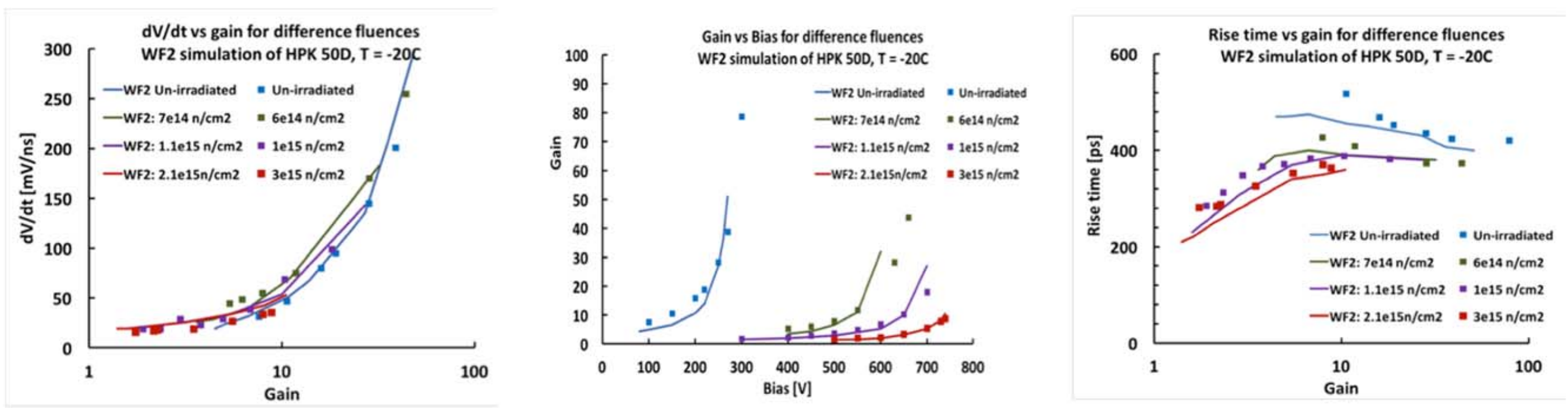

Figure 22 Comparison of Data and WF2 predictions of the slope dV/dt vs. gain (left), the gain vs bias (middle) and the rise time vs. gain (right).

## 8 CONCLUSIONS

We have performed a neutron radiation campaign on 50 µm thick UFSD produced by HPK up to a fluence of 6e15 n/cm$^2$. The sensors were operated at -20 $^o$C and -30 $^o$C to measure the leakage current, gain, timing resolution and pulse shapes as a function of bias voltage.

The following important points were explained in the paper:

- The bias voltage at breakdown increases from 300 V pre-irradiation to 750 V at the highest fluence.
- The gain at the breakdown voltage decreases from 70 (80) pre-irradiation to 2 (3) at the highest fluence at -20 $^o$C (-30 $^o$C).
- The fluence dependence of the gain supports the acceptor removal scenario.
- The operating bias voltage, the optimized CFD fraction, the rise time and the time resolution show a marked change at a fluence of 1e15 n/cm$^2$. This can be explained by the fact that the gain in the multiplication declines and a raised bias is needed to yield a sufficiently high electric field in the bulk, which then contributes to the avalanche process.
- For a CFD fraction optimized for each fluence and bias, the time resolution increases from 20 ps pre-irradiation to 40 ps after 1e15 n/cm$^2$ to 50 ps for 6e15 n/cm$^2$. The optimized CFD fraction stays constant at 10% to 15% from pre-irradiation up to 1e15 n/cm$^2$, and then increases to > 60%.
- Reducing the temperature from -20 $^o$C to -30 $^o$C improves the time resolution by 10% at 6e15 n/cm$^2$, but seems to have no influence on the resolution for 1e15 n/cm$^2$.
- Even though the gain for large fluences is low, of the order 2 - 3, the change in the pulse shape (especially the rise time) as a function of fluence is responsible for the good time resolution even after large neutron fluences.
- Lowering the optimal operating bias voltage by ~ 10% causes a reduced timing resolution of a few ps.
- The predictions of the Weightfield2 simulation package are in very good agreement with the experimental results.

## 9 ACKNOWLEDGEMENTS

We acknowledge the contribution to this paper by the HPK team of K. Yamamoto, S. Kamada, A. Ghassemi, K. Yamamura and the expert technical help by the SCIPP technical staff. Part of this work has been performed within the framework of the CERN RD50 Collaboration.

The work was supported by the United States Department of Energy, grant DE-FG02-04ER41286. Part of this work has been financed by the European Union’s Horizon 2020 Research and Innovation funding program, under Grant Agreement no. 654168 (AIDA-2020) and Grant Agreement no. 669529 (ERC UFSD669529), and by the Italian Ministero degli Affari Esteri and INFN Gruppo V.